\documentclass[a4paper,fleqn,usenatbib]{mnras}
\usepackage{hyperref}
\usepackage{newtxtext,newtxmath}

\usepackage[T1]{fontenc}
\usepackage{ae,aecompl}

\usepackage{graphicx}	
\usepackage{amsmath}	
\usepackage{amssymb}	
\usepackage{multicol}

\title[Matter of perspective]{The core - cusp problem: a matter of perspective}

\author[Genina et al.]{\newauthor Anna Genina$^{1}$\thanks{E-mail: anna.genina@durham.ac.uk}, Alejandro Ben\'{i}tez-Llambay$^1$\thanks{E-mail: alejandro.b.llambay@durham.ac.uk}, Carlos~S.~Frenk$^1$, Shaun Cole$^1$, \newauthor Azadeh Fattahi$^1$, Julio~F.~Navarro$^{2\thanks{Senior CIfAR Fellow}}$, Kyle~A.~Oman$^2$, Till
    Sawala$^{3}$ and Tom Theuns$^{1}$
\\
$^{1}$ Institute for Computational Cosmology, Department of Physics, Durham University, South Road, Durham, DH1 3LE, UK\\
$^{2}$  Department of Physics \& Astronomy, University of Victoria, Victoria, BC, V8P 5C2, Canada\\
$^{3}$ Department of Physics, University of Helsinki, Gustaf
    H\"allstr\"omin katu 2a,
    FI-00014 Helsinki, Finland
}

\date{Accepted XXX. Received YYY; in original form ZZZ}

\pubyear{2017}

\begin{document}
\label{firstpage}
\pagerange{\pageref{firstpage}--\pageref{lastpage}}
\maketitle


\begin{abstract}

  \noindent The existence of two kinematically and chemically distinct
  stellar subpopulations in the Sculptor and Fornax dwarf galaxies
  offers the opportunity to constrain the density profile of their
  matter haloes by measuring the mass contained within the
  well-separated half-light radii of the two metallicity
  subpopulations. Walker and Pe\~{n}arrubia have used this approach to
  argue that data for these galaxies are consistent with
  constant-density `cores' in their inner regions and rule out `cuspy'
  Navarro--Frenk--White (NFW) profiles with high statistical
  significance, particularly in the case of Sculptor. We test the
  validity of these claims using dwarf galaxies in the APOSTLE (A Project Of Simulating The Local Environment)
  $\Lambda$ cold dark matter cosmological hydrodynamic simulations of analogues of
  the Local Group. These galaxies all have NFW dark matter density
  profiles and a subset of them develop two distinct metallicity
  subpopulations reminiscent of Sculptor and Fornax. We apply a method
  analogous to that of Walker and Pe\~{n}arrubia to a sample of 50
  simulated dwarfs and find that this procedure often leads to a
  statistically significant detection of a core in the profile when in
  reality there is a cusp. Although multiple factors contribute to
  these failures, the main cause is a violation of the assumption of
  spherical symmetry upon which the mass estimators are based. The
  stellar populations of the simulated dwarfs tend to be significantly
  elongated and, in several cases, the two metallicity populations
  have different asphericity and are misaligned. As a result, a wide
  range of slopes of the density profile are inferred depending on the
  angle from which the galaxy is viewed.

\end{abstract}



\section{Introduction} \label{sec1}

One of the fundamental predictions of the $\Lambda$ cold dark matter ($\Lambda \mathrm{CDM}$) model
of cosmogony is that dark matter assembles into haloes that, in the
absence of baryon effects, develop steeply rising inner radial density
profiles, or cusps. This important result was obtained from $N$-body
simulations which showed that the density distribution of a dark
matter halo of any mass is well fit by a Navarro--Frenk--White profile
\citep[NFW;][]{navarro1996b,nfw} independently of initial conditions
and cosmological parameters.

The inner slope of the NFW profile follows $\rho \propto r^{-1}$.  In
contrast, measurements of galaxy rotation curves and dynamical models
of dwarf spheroidal galaxies are often claimed to require shallower
density profile slopes that are consistent with a constant-density
core at the centre, $\rho \propto r^{0}$
\citep[e.g.][]{Moorecore,flores,battaglia,walk,amorisco,agnello,adamsgas,littlethings}. This
disagreement between observations and simulations has become known as
the \textit{core-cusp problem}.

In order to resolve this discrepancy, a number of mechanisms involving
baryons, which could transform cusps into cores, have been
proposed. For example, cores may be created when baryons, after slowly
condensing at the centre of a halo, are suddenly expelled by
supernova feedback, either in a single event
\citep{navarro1996b,readgilmore} or through repeated episodes of star
formation \citep{pontzen,brooks2014}.  Alternatively, energy could be
transferred to the outer halo by clumps infalling due to dynamical
friction \citep{sanchez,maschenko,colecore,elzant,delpopolo}, or
through resonant effects induced by a central stellar
bar \citep{weinberg_katz2002}.

Dwarf spheroidal galaxies are promising objects to test ideas
about the inner structure of dark matter haloes. These galaxies are
strongly dark-matter dominated \citep{pryor} and, although they are
faint, some are sufficiently close-by that their stellar populations
can be resolved.  Much effort has therefore been invested in trying to
infer their halo profiles. A large body of work is concerned with
field galaxies with measurable H\,{\sc i} velocity fields; many such
studies claim robust detections of central cores
\citep[e.g.][]{ohthings,kuzio,adamsgas}. However, a recent study
\citep{kylecurves}, based on the same APOSTLE (A Project Of Simulating The Local Environment) simulations that we will
analyse here, has revealed the presence of systematic effects in even
the most detailed analyses of spatially resolved kinematics, casting
doubt on claims that cores are present in those galaxies. Similar
conclusions were reached by \citet{fatalattraction}.

The kinematics of resolved stars in nearby galaxies offer an
alternative to the kinematics of H\,{\sc i} gas as a probe of the
density structure of haloes. The detection of cores in several dwarf
satellites of the Milky Way has been claimed on the basis of simple
Jeans analyses \citep[e.g.][]{gilmore2007}, but the more general
analysis by \citet{strigari2010} has shown that current data are, in
fact, unable to distinguish between cores and cusps in the Milky Way
satellites. Some satellites of the Milky Way and Andromeda exhibit
metallicity gradients: they have a centrally concentrated metal-rich
population and a more extended, and kinematically hotter metal-poor
population \citep{tolstoy,battaglia,battagliasextans}. The origin of
these systems is unknown but major mergers \citep{mergers},
reaccretion of gas \citep{tolstoy,battaglia06} or effects due to
reionization \citep{photoion} have been proposed as possible origins
of metallicity gradients.

The presence of two kinematically and spatially distinct metallicity
components can be used to set constraints on the inner density profile
of the common halo in which they move. \citet{battaglia} identified a
metal-rich ($\mathrm{ \left[Fe/H \right] > -1.5}$) and a metal-poor
($\mathrm{ \left[Fe/H \right] <-1.7}$) population in the Sculptor
dwarf spheroidal and, using Jeans modelling, found that a wide range
of profiles are consistent with the data, from a pseudo-isothermal
sphere ($\rho \propto r^{0}$ at the centre) to an NFW
profile. \citet{amorisco} pointed out that some of those models are
unphysical and, fitting the Sculptor data to a particular phase-space
distribution function, found that while a profile with a core is
preferred by their $\chi^2$ fits, an NFW profile is also allowed by
the data.  Using more general phase-space distribution functions,
\citet{strigari} also showed that the two metallicity subpopulations in
Sculptor are consistent with an NFW profile. A similar conclusion,
using Schwarzschild modelling, was reached by \citet{breddels} who
found that a core profile is also allowed, while a cusp in Sculptor
was found to be favoured by an analysis based on the fourth-order
virial theorem by \citet{richardson}.

\citet{walk} took this idea further and developed a statistical
methodology to distinguish the two metallicity subpopulations in
Sculptor and Fornax. Making use of the interesting result of
\cite{wolf} and \cite{walkerest} that the mass of a spherical stellar
system in equilibrium can be robustly estimated at the half-mass
radius of the system, they developed the method discussed in this
paper and concluded that both Sculptor and Fornax have central cores,
with Sculptor, in particular, ruling out an NFW profile at high
statistical significance.  Their method is based on estimating the
total mass contained within the half-light radii of the metal-rich and
metal-poor subpopulations, thus constraining the slope of the dark
matter density profile. \cite{wolf} and \cite{walkerest} have argued
that the mass within a characteristic radius of a collisionless
spherical system in dynamical equilibrium is well constrained by the
velocity dispersion and average radial distribution of a population of
star tracers, for a variety of stellar density and constant velocity
anisotropy profiles.

Specifically, \cite{wolf} showed that the mass is best constrained at
radius, $r_3$, where the logarithmic slope of the stellar number
density, $\rm{d}\log\nu/ \rm{d}\log r=-3$, which is close to the
deprojected half-light radius ($\frac{4}{3}R_{\mathrm{e}}$, where
$R_{\mathrm{e}}$ is the projected half-light radius) for a
range of stellar distributions. Their estimator is:
\begin{equation}
\label{wolfeq}
M \left( <\frac{4}{3}R_{e} \right) = 4G^{-1}\langle \sigma_{\mathrm{LOS}}\rangle^2 R_{e} \,,
\end{equation}
where $\langle \sigma_{\mathrm{LOS}}\rangle$ is the
\textit{luminosity-averaged} line-of-sight velocity dispersion.
Similarly, \cite{walkerest} propose:
\begin{equation}
\label{walkereq}
M\left( <R_e \right) = \frac{5}{2}G^{-1} \langle \sigma_{\mathrm{LOS}}\rangle^2 R_{e} \,.
\end{equation}

\citet{walk} used a likelihood method to separate samples of stars in
Sculptor and Fornax into two metallicity subcomponents and applied these
mass estimators to each of them. For an object with mass density $\rho\propto
r^{-\gamma}$ the enclosed mass is $M(r)\propto \frac{1}{3 - \gamma}
r^{3-\gamma}$. One can then define the asymptotic logarithmic mass slope
as:
\begin{equation}
\label{gammadef}
\Gamma = \lim_{r \to 0} \left[ \frac{\rm{d} \log M}{\rm{d} \log r} \right] = 3 - \gamma
\end{equation}
For an NFW profile the asymptotic inner slope is $\gamma$ = 1, so
$\Gamma=2$, while for a core with $\gamma$ = 0, $\Gamma= 3$. In the
case of a galaxy with two segregated subpopulations, the two
half-light radii will be located away from the centre and thus $\Gamma
= \Delta \log M/ \Delta \log r$ is a measure of the (steeper) density
slope further out \citep{walk}. Assuming that the mass is given by the
estimators above,
\begin{equation}
\label{taylor}
\Gamma \approx 1 + \frac{\log \left( \sigma_2^2 / \sigma_1^2 \right)}{\log \left( r_2/r_1 \right)} \,,
\end{equation}
where $\sigma_i$ are the line-of-sight velocity dispersions and $r_i$ are
the half-light radii. \citet{walk} derived values of $\Gamma$ for
Sculptor and Fornax which exclude an NFW cusp at 99 and 96 per cent
confidence levels, respectively, instead favouring a core. They argue
that this conclusion is conservative because, if anything, the mass
is likely to be overestimated for the central subpopulation.

In a recent paper \citet{campbell} tested the accuracy of the mass
estimators by applying them to galaxies in the APOSTLE hydrodynamic
cosmological simulations of Local Group analogues (for which the true
mass is known; see Section \ref{sim}). They report little bias in the
median mass estimates but a scatter of 25 and 23 per cent for the
\citet{walkerest} and \citet{wolf} estimators respectively, which are
much larger than the values inferred by these authors from simulations
of spherical systems in dynamical equilibrium. \citet{campbell} find
that a major contribution to the scatter comes from deviations from
spherical symmetry which are quite common in simulated galaxies.
Subsequently, \citet{fireestimator} have generally confirmed the main
conclusions of \citet{campbell} regarding the scatter in the estimator
from 12 dwarf galaxy analogues in the FIRE hydrodynamic
simulations.

The effect of triaxality on $\Gamma$ has been investigated by
\citet{triaxality}, who tagged dark matter particles as stars in the
(triaxial) dark matter haloes of the AQUARIUS simulations
\citep{aquarius}. These authors find that an anti-correlation between
the measured half-light radius and the projected velocity dispersion
acts to keep the mass estimate approximately constant, causing little
variation in the derived value of $\Gamma$. However, by construction,
the dark matter and the stars in their analysis have strongly
correlated shapes \citep{laporteconstruction} which can introduce a
systematic effect.

\citet{orientation} carried out idealized $N$-body simulations of the
evolution of dwarf spheroidal galaxies in the gravitational potential
of a Milky Way-like host. They introduced two spatially segregated disc
subpopulations which evolve as the dwarf orbits in the halo of its
host. They find that $\Gamma$ may be over- or underestimated,
depending on the line of sight. In particular, observations along the
major axis of the dwarf tend to overestimate the mass and
$\Gamma$. These simulations do not take into account dark matter halo
triaxality, hydrodynamics, star formation or feedback processes.

\citet{campbell} focused on the accuracy of the mass estimators
applied to the stellar population of the dwarf galaxies in APOSTLE as
a whole. Many of these galaxies, however, turn out to have two (or
more) distinct metallicity subpopulations, analogous to those in Sculptor or Fornax. This offers the possibility of testing the
validity of the conclusions of \citet{walk} using realistic dwarf
galaxies formed in state-of-the-art cosmological simulations. This is
the goal of this paper.  In Section~\ref{sim} we outline the selection
of our simulated galaxy sample and perform an analogous analysis to
that of \citet{walk}. In Section~\ref{cases} we examine in detail a
selection of case studies. We summarize our conclusions in Section
\ref{conc}.

\begin{figure}

		\includegraphics[width=0.9\columnwidth]{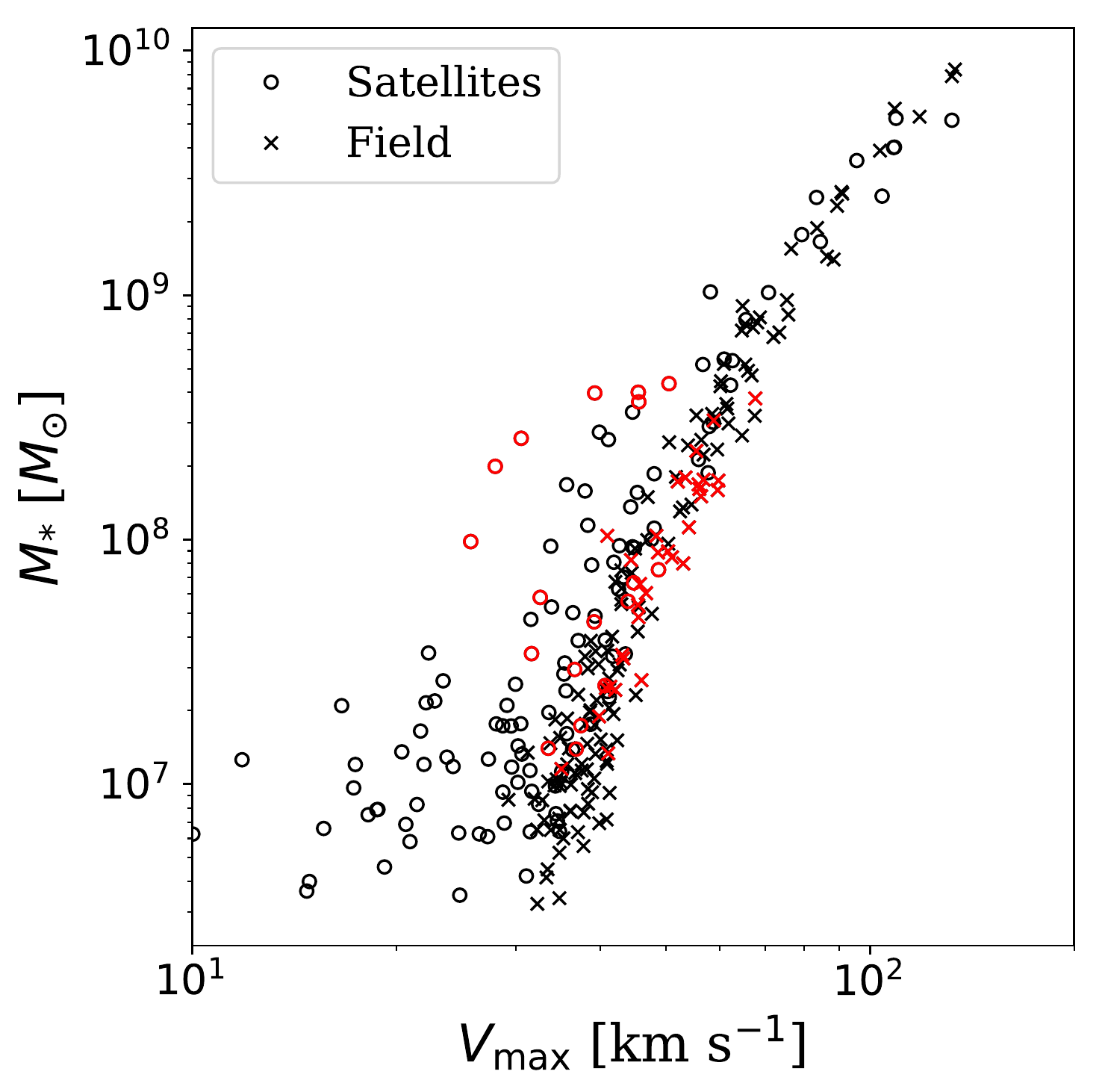}

		\caption{ Stellar mass $M_{*}$ as a function of maximum circular velocity
                  $V_\mathrm{max}$ for satellite (circles) and field
                  (crosses) galaxies in the five high-resolution
                  volumes of APOSTLE. Galaxies matching the criteria
                  described in Section \ref{splitting} are shown in
                  red. The sample is limited to galaxies with at least
                  1000 stellar particles to ensure adequate statistics. }

		\label{m200}
	\end{figure}

\section{Simulations and methods} \label{sim}

\begin{table}
\centering
\caption{Dark matter, gas and the ranges of stellar particle masses for the five L1 volumes of {\sc{APOSTLE}} used in this work. The
  gravitational softening at $z$ = 0 is 134 pc. }
\label{table1}
\begin{tabular}{cccc}
Volume & $\mathrm{m}_{\mathrm{DM}}\mathrm{h}^{-1}$ [M$_{\odot}$]  & $\mathrm{m}_{\mathrm{gas}}\mathrm{h}^{-1}$ [M$_{\odot}$] & $\mathrm{m}_{\mathrm{star}}\mathrm{h}^{-1}$ [M$_{\odot}$] \\
\hline
AP-1 &  3.5 $\times 10^{4} $ & 7.0 $\times 10^{3} $ & 0.4-1.4 $\times 10^{4}$ \\
AP-4 &  1.7 $\times 10^{4} $ & 3.5 $\times 10^{3} $ & 0.2-1.0 $\times 10^{4}$ \\
AP-6 &  3.7 $\times 10^{4} $ & 7.5 $\times 10^{3} $ & 0.4-2.5 $\times 10^{4}$ \\
AP-10& 3.6 $\times 10^{4} $ & 7.2 $\times 10^{3}  $ & 0.4-1.2 $\times 10^{4}$
\\
AP-11& 3.5 $\times 10^{4} $ & 7.1 $\times 10^{3} $ & 0.4-1.6 $\times 10^{4}$
\end{tabular}
\end{table}

\subsection{APOSTLE simulations}

APOSTLE consists of
a suite of zoom-in hydrodynamical simulations of analogues of the
Local Group environment \citep{fattahi,sawalapuzzles}. The regions
were selected for resimulation from the 100 Mpc on a side
cosmological $N$-body simulation DOVE \citep{dove}. The Milky Way - Andromeda analogues were chosen
based on the galaxy pair separations, total mass, relative velocities,
recession velocities of the outer Local Group members and consistency
with the environment surrounding the Local Group. WMAP-7 cosmological
parameters are assumed: density parameters, $\Omega_\mathrm{m}$ =
0.272, $\Omega_\mathrm{b}$ = 0.0455 and $\Omega_{\mathrm{\Lambda}}$ =
0.728; reduced Hubble constant $h$ = 0.704;  spectral index $n_s$ =
0.967 and power spectrum normalization,  $\sigma_8$ = 0.81
\citep{wmap7}. An ionizing background is switched on instantaneously at
$z$=11.5.

The regions were resimulated using the {\sc{eagle}} code, an improved version
of the $N$-body/Smooth Particle Hydrodynamics (SPH) code P-{\sc{gadget}}-3
\citep{eagle1,eagle2,gadget}, including subgrid prescriptions for
supernovae and AGN feedback \citep{anarchy,agnbooth}, gas cooling and
heating \citep{gascooling}, reionization, star formation and metal
enrichment \citep{starformation,starformation2,enrichment} and black hole
formation and mergers \citep{accretionmergers}.  The Tree-PM scheme of
P-{\sc{gadget}}-3 is used to compute the gravitational acceleration and the
ANARCHY SPH scheme \citep{anarchy,anarchy2}, based on the
pressure-entropy formalism of \citet{hopkins}, is used to compute
hydrodynamical forces.

The APOSTLE suite consists of 12 volumes simulated at low and
medium resolution (L2 and L3). Five of these were also resimulated at high
resolution (L1). In this work we will only consider galaxies within the
high-resolution volumes. The gas, dark matter and stellar particle masses for
each of these may be found in Table \ref{table1}.

\subsection{Galaxy sample}
\label{splitting}
Haloes in the simulations are identified using the
`friends-of-friends' (FOF) algorithm with linking length of 0.2
times the mean particle separation \citep{fof}. The SUBFIND algorithm
is then used to identify gravitationally bound substructures within
them \citep{subfind}. We define the host and subhalo centres as the
centre of their potential (the position of the particle with the most
negative potential energy).

Subhaloes bound to the main halo of a group are defined here as
`satellites'; other galaxies in the volume are labelled as `field'
galaxies. When computing the stellar mass of a subhalo, we
include all particles located within 0.15 of the
  virial radius, $R_{200}$, for field galaxies and particles located
  within the tidal radius\footnote{We define the tidal radius of a
    subhalo as a distance from subhalo centre where the mean enclosed
    density is equal to that of the host halo up to that distance for satellite galaxies.
  for satellite galaxies}. We limit the sample of satellites and field
galaxies to those with a minimum of 1000 stellar particles
[corresponding to a stellar mass of the order of $(10^6-10^7)$ $M_{\odot}$]
to ensure reasonable statistics and good resolution within the
half-mass\footnote{We use the term \textit{half-mass radius} to refer
  to the radius enclosing half the stellar mass as measured directly
  from the simulations. } radius. The stellar mass as a function of
the maximum circular velocity, $\rm{V}_{\mathrm{max}}$, of galaxies in
the five high resolution volumes is shown in Fig.~\ref{m200}.

\begin{figure}

\includegraphics[width=\columnwidth]{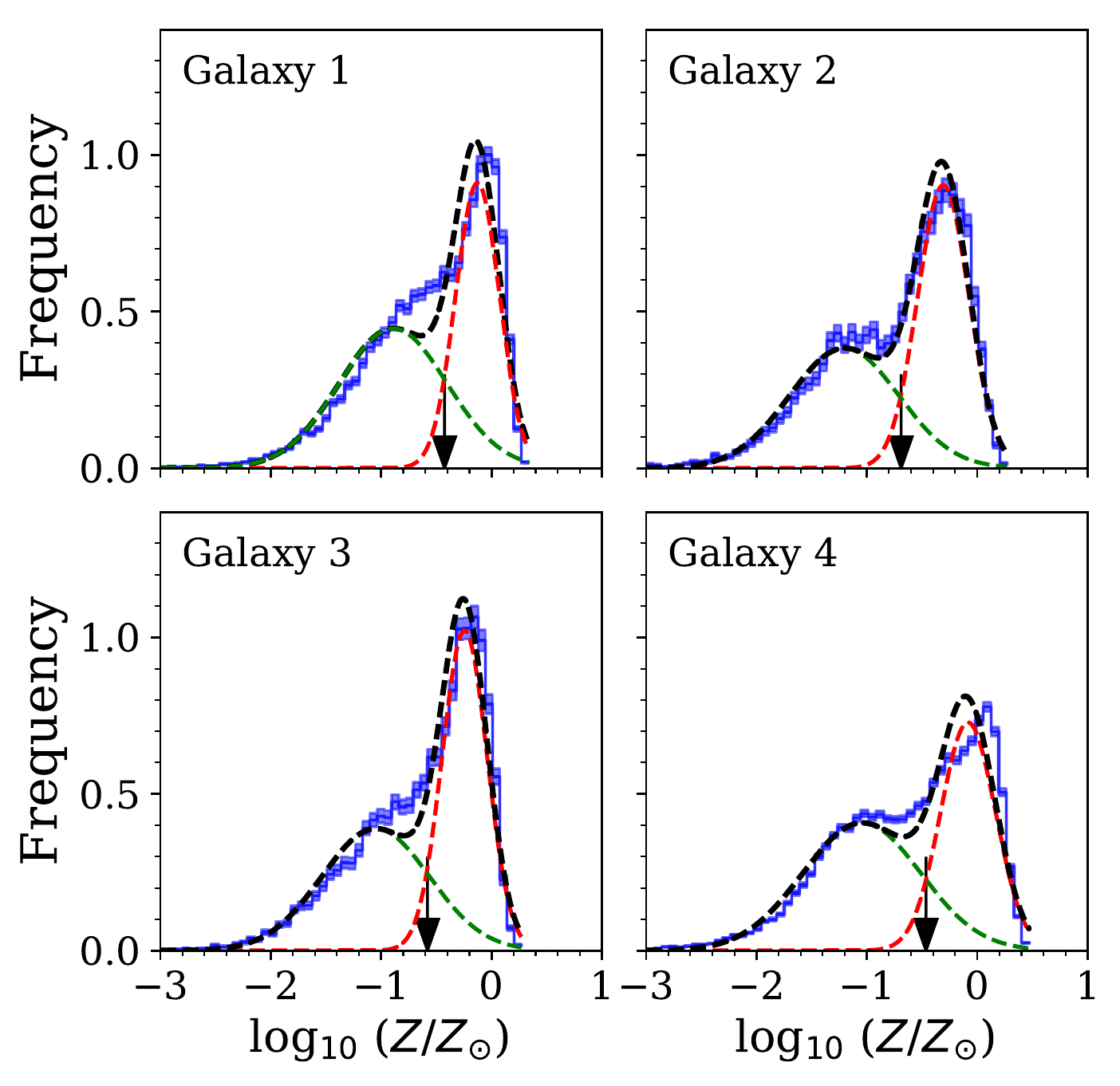}

\caption{Metallicity histograms of the four galaxies that we use as
  case studies in this work. The blue curve represents the metallicity
  distribution and associated Poisson errors. The dashed black line
  shows the two-Gaussian fit. The dashed red and green lines show the
  individual Gaussians corresponding to the metal-rich and metal-poor
  subpopulations, respectively. The arrows show the metallicity at
  which the population is split. We have assumed a value of the
  solar metallicity of $Z_{\odot}$ = 0.0127.}

\label{hist}
\end{figure}

In order to identify particles belonging to each stellar
subpopulation, for every individual satellite and field galaxy we
model the subcomponents using Gaussian Mixture Modelling (GMM) whereby
the total metallicity distribution, $p(\log_{10}
\rm{Z}/\rm{Z}_{\odot})$, is fitted with a combination of two
Gaussian probability density functions\footnote{As a measure of metallicity we use $\log_{10}
  Z/Z_{\odot}$, the logarithm of the abundance of elements other than
  hydrogen and helium. Stellar particles in APOSTLE are spawned
  probabilistically, with daughter particles inheriting smoothed metal
  abundances from their parent. For details see
  \citet{metal2,metal1}.} \citep{gmm}. Five parameters
are fit altogether ($w_1$, $\mu_1$, $\mu_2$, $\sigma_1$, $\sigma_2$),
with $w_1$ being the relative weight of one of the subpopulations,
$\mu_i$ the mean metallicity and $\sigma_i$ the metallicity
dispersion. We then assign each particle to a subpopulation if its
probability of being in that subpopulation is $p(i) > 0.5$. Effectively,
the population is rigidly split at the value of metallicity where the
two Gaussians cross. The subpopulation with a higher value of $\mu_i$ is
denoted as metal-rich and that with the lower $\mu_i$ as metal-poor.
A sharp cut in metallicity gives rise to some kinematic mixing
of the two subpopulations. We have verified that mixing  has only a
minor effect on the main results of this paper regarding the slope of
the halo density profile (see Section~\ref{mixing}).

It is important to note, first, that the metallicity distributions
of the two subpopulations will not necessarily be Gaussian, but will
depend on the specifics of the history of star formation, accretion
and mergers. We choose Gaussian probability densities for
simplicity. Secondly, cases may exist, where indeed more than two
subpopulations are present. These objects would be of great interest
for future work due to the possibility of constraining the inner
density slope at two or more locations.

In principle, any probability distribution will be better fitted with
Gaussian mixtures as the number of fitting parameters is increased. We
therefore calculate the Akaike Information Criterion (AIC) corrected
for finite sample \citep{Akaike}:
\begin{equation}
\label{aic}
\mathrm{AIC} = 2k + \chi^2 + \frac{2k(k+1)}{n-k-1} \, ,
\end{equation}
where $k$ is the number of fitted parameters, $n$ is the number of
data points and $\chi^2$ is the chi-squared fit of our model to the
data. We take the histogram errors to be Poisson-distributed. The
first and third terms in equation~(\ref{aic}) represent the penalty on the
number of free parameters in the model such that the difference
between the AIC values for alternative models is indicative of the
information gained by including extra parameters. We find the AIC for a model with a
single Gaussian and a model with a mixture of two Gaussians for
each galaxy in our sample. We then remove objects where the AIC for a single Gaussian is
smaller than that for a mixture of two, as well as those where both
models provide a poor fit. A total of 46 per cent of all galaxies
satisfy these criteria. The metallicity histograms and subpopulation
models of specific objects that we will discuss in particular detail
later are shown in Fig.~\ref{hist}.

A simple split into two metallicity subcomponents does not guarantee
that they will be spatially segregated. We remove from our sample the
objects for which the separation between the two half-light radii is
so small as to inflate $\Gamma$ artificially, as $\Delta \log_{10} r$
approaches zero (see Fig.~\ref{appendix:blowup}). We therefore discard
all objects for which the logarithm of the ratio of 3D half-mass
radii, $\log_{10}(r_2/r_1)$ $<$ 0.06. This condition removes a further
26 per cent of our original sample. For the remaining galaxies, we
check that the metal-rich and metal-poor radii are well resolved as
judged by the convergence radius defined by \citet{power}, at which
the collisional relaxation time is approximately equal to the age of
the Universe, ensuring that both radii are larger than this
value. Overall, of all objects with over 1000 stellar particles in the
five high-resolution volumes of APOSTLE, 18 per cent (50 objects)
survive our selection criteria. The selected objects
  are shown in red in Fig. \ref{m200}; they have stellar masses of the
  order of 10$^7$ - 10$^8$ $M_{\odot}$. We find that the fraction of
  stellar particles assigned to the metal-poor subpopulation ranges between 0.15
  and 0.6, consistently with the results of \citet{mergers}; the
  metal-rich stellar particles are typically the dominant subcomponent.

\begin{figure}
\centering
\includegraphics[width=\columnwidth]{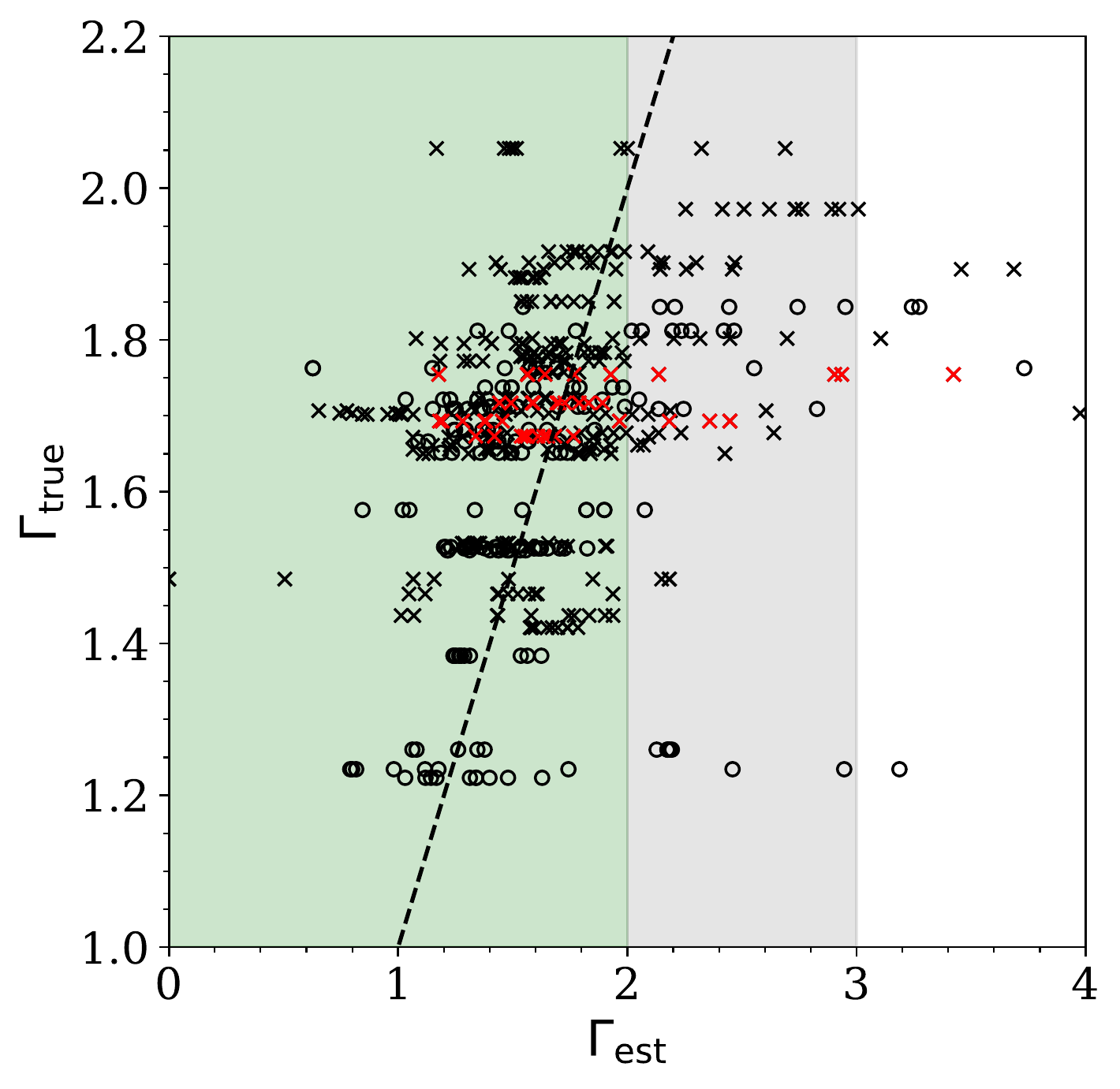}
\caption{The slope of the logarithmic mass distribution,
  $\Gamma_{\mathrm{est}}$, obtained by applying the \citet{walkerest} estimator to
  simulated galaxies in our sample viewed from 10 random directions,
  plotted against the true slope at the projected 3D half-mass radius,
  $\Gamma_{\mathrm{true}}$. Field galaxies are shown as black crosses
  and satellites as black circles. Marked in red are the four
  particular cases that will be further discussed in detail in the
  next section. The green vertical band shows typical values of
  $\Gamma_{\mathrm{est}}$ for a cusp, ($\Gamma = 2$ corresponds to an
  NFW cusp); the grey band shows values of $\Gamma_{\mathrm{est}}$
  typical for a core. The dashed line is the one-to-one locus. The
  slopes obtained from the estimator are slightly underestimated  but
  exhibit significant scatter as the line of sight varies.} 
\label{LOS}
\end{figure}

\begin{figure*}
\centering
\includegraphics[width=2.\columnwidth]{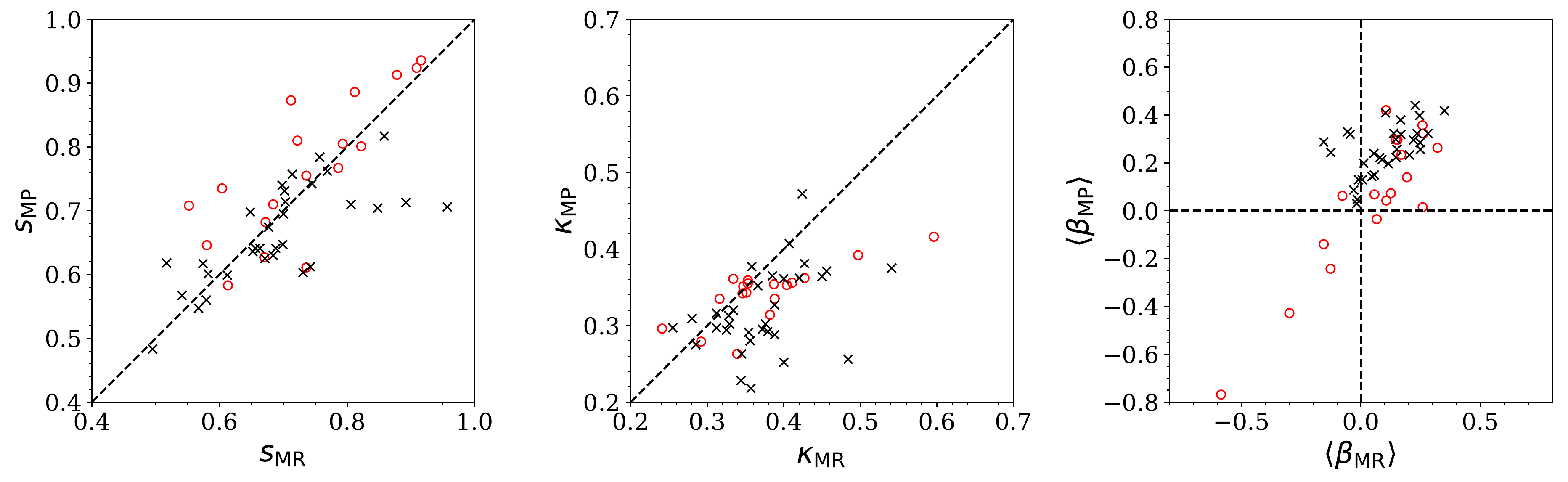}
\caption{\textit{Left:} the correlation of the sphericities of the
  metal-poor and the metal-rich subpopulations for satellites (red
  circles) and field galaxies (black crosses) in our sample. The
  dashed line indicates the one-to-one relation. \textit{Middle:}
  $\kappa_{\mathrm{rot}}$ of the metal-poor and metal-rich
  subpopulations. \textit{Right:} the correlation between the average velocity
  anisotropy of the metal-poor and metal-rich subpopulations. The
  dashed lines indicate isotropy, $\beta$=0. }
\label{properties}
\end{figure*}

\subsection{A test of the Walker-Pe\~{n}arrubia prescription}
\label{sec:WP}

We now carry out a straightforward test of the accuracy of the
logarithmic mass slopes obtained following the \citet{walk}
prescription. We generate 10 random lines of sight distributed
uniformly on the surface of a sphere. For each line of sight, we
obtain 1000 bootstrap stellar particle samples for each galaxy, with
replacement and, for each sample, we calculate projected half-mass radii,
$R_{\mathrm{e}}$, directly as the projected radius within which half
the total stellar mass is contained\footnote{According to
  \citet{fireestimator} this method of estimating $R_{\mathrm{e}}$
  results in the bias of $\sim$ 0.9 in the estimated mass seen in the
  12 FIRE simulations. However, using a much larger sample of
  galaxies in APOSTLE, \citet{campbell} found that the mass estimate
  is, in fact, unbiased. We choose to calculate the projected
  half-mass radii as in \citet{campbell}.}. We then calculate the
mass-averaged line-of-sight velocity dispersion as:
\begin{equation}
 \sigma_{\mathrm{LOS}}^2 =\frac{\sum\left(v_i - \bar{v}\right)^2 m_i}{\sum m_i} \, ,
\end{equation}
where $m_i$ is the mass of each star particle, $v_{i}$ is the
velocity of the particle in projection and $\bar{v}$ is the
mean velocity.

Inserting our measured values of $R_{\mathrm{e}}$ and $\langle
\sigma_{\mathrm{LOS}} \rangle$ in the \citet{walkerest} estimator (equation~\ref{walkereq}) we obtain the estimated mass within $R_{\mathrm{e}}$ of
each subpopulation.  For each galaxy, we repeat this calculation for
the 1000 bootstrap resamplings and for 10 random directions.  In
Fig.~\ref{LOS} we plot $\Gamma_{\mathrm{true}}$, the
slope of the line joining the logarithm of the actual mass within the
true projected 3D half-mass radius of each subpopulation\footnote{We take the true projected 3D half-mass radius of an object to be 3/4 of the 3D half-mass radius measured from the simulation.}, as a function of the median values of $\Gamma_{\mathrm{est}}$,
the slope of the line joining the logarithm of the estimated mass within the measured $R_{\mathrm{e}}$ of each subpopulation.

It is clear that the estimated mass slopes tend to be underestimated on
average, consistent with findings of \citet{walk}, and thus the inferred slopes of the density profiles tend to
be cuspier than the true values.  (Recall that
$\Gamma=2$ corresponds to an NFW cusp, while
$\Gamma=3$ corresponds to a core.)  The distribution
is asymmetric and exhibits large scatter towards higher values of
$\Gamma_{\mathrm{est}}$, with some objects reaching
$\Gamma_{\mathrm{est}} \geq 3$. This bias reflects biases in the
measurements of $R_{\mathrm{e}}$ and $\langle \sigma_{\mathrm{LOS}}
\rangle$ as the galaxy is seen from different observer positions.  We
now investigate why the measured half-mass radii and velocity
dispersions vary with the viewing direction.

\subsection{Dynamical properties of the simulated galaxies}
\label{props}
\citet{campbell} identified asphericity, rotation and velocity
anisotropy as the key properties that can introduce uncertainty in
mass measurements based on stellar kinematics.  We now quantify these
properties for \textit{each} metallicity subpopulation within each galaxy in our sample and examine the extent to which the properties of
the two subpopulations are correlated with each other\footnote{As discussed in Section~\ref{numerical}, we have checked
    that the two subpopulations of the galaxies in our sample have a
    sufficiently large number of stellar particles for the properties
    of interest to be numerically converged.}.

\subsubsection{Sphericity}
Here we define the centre of each stellar subpopulation as its centre of mass. The
shape of the system is characterized by the reduced inertia tensor
\citep{sphericity}:
\begin{equation}
\label{spherecalc}
I_{ij} = \frac{1}{M_{*,\mathrm{sub}}} \sum_{n}m_{n}\frac{r_{n,i} r_{n,j}}{r_{n}^2} \,,
\end{equation}
where $M_{*,\mathrm{sub}}$ is the stellar mass of the subpopulation;
$m_{n}$ the mass of star particle $n$; $r_{n,i}$ and $r_{n,j}$ are the
coordinates of particle $n$ from the centre of the galaxy in directions
$i$ and $j$. The normalization $r_n^2$ ensures that only the angular
distribution is taken into account, so that the shape is not unduly
affected by distant particles. The eigenvectors of the inertia tensor
correspond to the axes of the fitted ellipsoid and the eigenvalues,
$a^2\geq b^2 \geq c^2$ to squares of axis lengths. We define the sphericity
$s=c/a$; $s=1$ corresponds to a sphere.

The sphericities of the metal-rich and metal-poor subpopulations in
our sample are plotted against each other in the left panel of
Fig.~\ref{properties}. The two are positively correlated, albeit with
significant scatter caused by one of the subpopulations in certain
objects being appreciably more spherical than the other. These cases
are of particular interest in this study. Also note that the
satellites tend to be less aspherical than the field
galaxies. This is likely due to the effects of tidal stripping as
discussed in detail in the work of \citet{barber}.

\subsubsection{Rotation}

We quantify the degree to which each subpopulation is supported by
rotation by computing $\kappa_{\mathrm{rot}}$, the fraction of
kinetic energy invested in rotational motion \citep{kappa}:
\begin{equation}
\kappa_{\mathrm{rot}} = \frac{1}{K_{*,\mathrm{sub}}} \sum_n \frac{m_n}{2} \left( \frac{j_{z,n}}{R_{xy,n}} \right)^2,
\label{kapparoteq}
\end{equation}
where $K_{*,\mathrm{sub}}$ is the stellar kinetic energy of the
subpopulation, $m_n$ the mass of star particle $n$, $j_{z,n}$ the
component of the specific angular momentum of the particle in the
direction of the total angular momentum and $R_{xy,n}$ the
distance of the particle from the angular momentum axis. Objects with
$\kappa_{\mathrm{rot}} > 0.5$ are considered to be primarily
rotation-dominated, while objects with $\kappa_{\mathrm{rot}}$ < 0.5
are considered to be primarily dispersion-dominated.

The values of $\kappa_{\mathrm{rot}}$ for subpopulations of galaxies
in our sample are shown in the middle panel of
Fig.~\ref{properties}. Our selected objects are generally
dispersion-dominated and a strong bias exists towards higher
$\kappa_{\mathrm{rot}}$ in the metal-rich subpopulation compared to
the metal-poor. All simulated galaxies
considered in this work have $\kappa_{\mathrm{rot}} <$ 0.5 for the
galaxy as a \textit{whole}.

\subsubsection{Velocity anisotropy}

The velocity anisotropy is defined as
$\beta(r)$=$1$-$\sigma^2_t$/2$\sigma^2_r$, where $\sigma_r$ is the
radial velocity dispersion and $\sigma_t$ the tangential velocity
dispersion including the contributions from azimuthal and polar
directions.  We construct velocity anisotropy profiles by calculating
$\sigma_r$ and $\sigma_t$ for the 32 nearest neighbours of each
star particle.

The right panel of Fig.~\ref{properties} shows the average velocity
anisotropy, $\langle \beta \rangle$, for each metallicity
subpopulation (the average of local anisotropy of each particle). The majority of the galaxies in our sample tend to have
radially biased stellar velocity distributions and the anisotropies
are generally correlated in the two metallcity subpopulations. Yet,
cases exist where $\langle \beta \rangle$ is radially biased for one
subpopulation and tangentially biased or isotropic for the other. As
we shall see, such discrepancies affect estimates of $\Gamma$.

\section{The effects of projection: four case studies}
\label{cases}
In Fig.~\ref{LOS} we saw that a procedure analogous to
  that implemented by \citet{walk} in their analysis of the kinematics
  of Sculptor and Fornax dwarfs generally underestimates the
  logarithmic mass slopes $\Gamma$, albeit with a large scatter towards higher
  values, which would correspond to shallower inner density slopes. We now
investigate the factors that affect the accuracy of the procedure. We
first examine in detail four illustrative examples and in
Section~\ref{general} we collect the results for our sample of 50
galaxies. We recall that the dark matter density profiles of all the
galaxies in our sample are well described by an NFW profile.

\begin{figure*}
\label{props}
\begin{multicols}{2}
    \includegraphics[width=\linewidth]{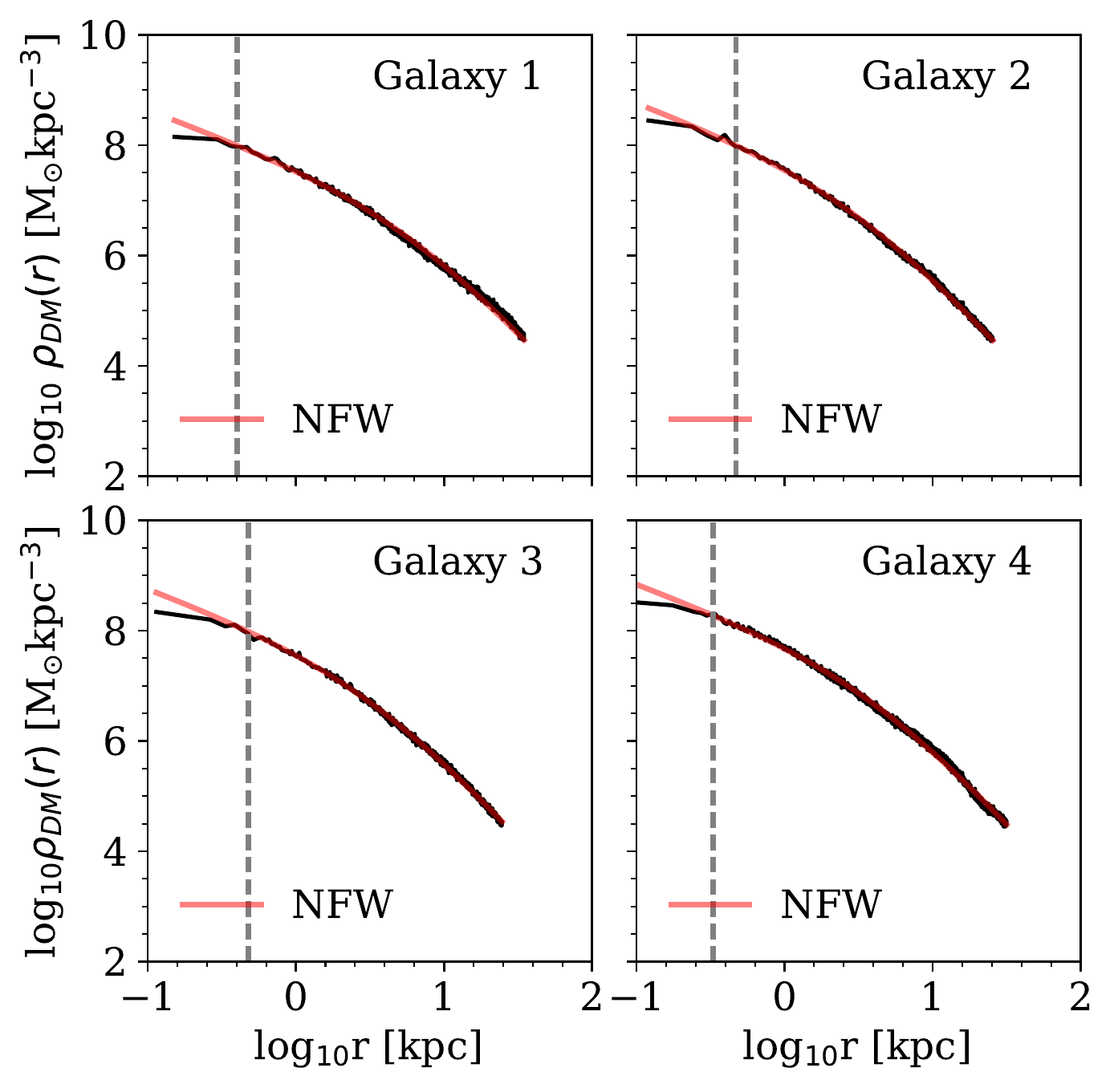}\par
    \includegraphics[width=\linewidth]{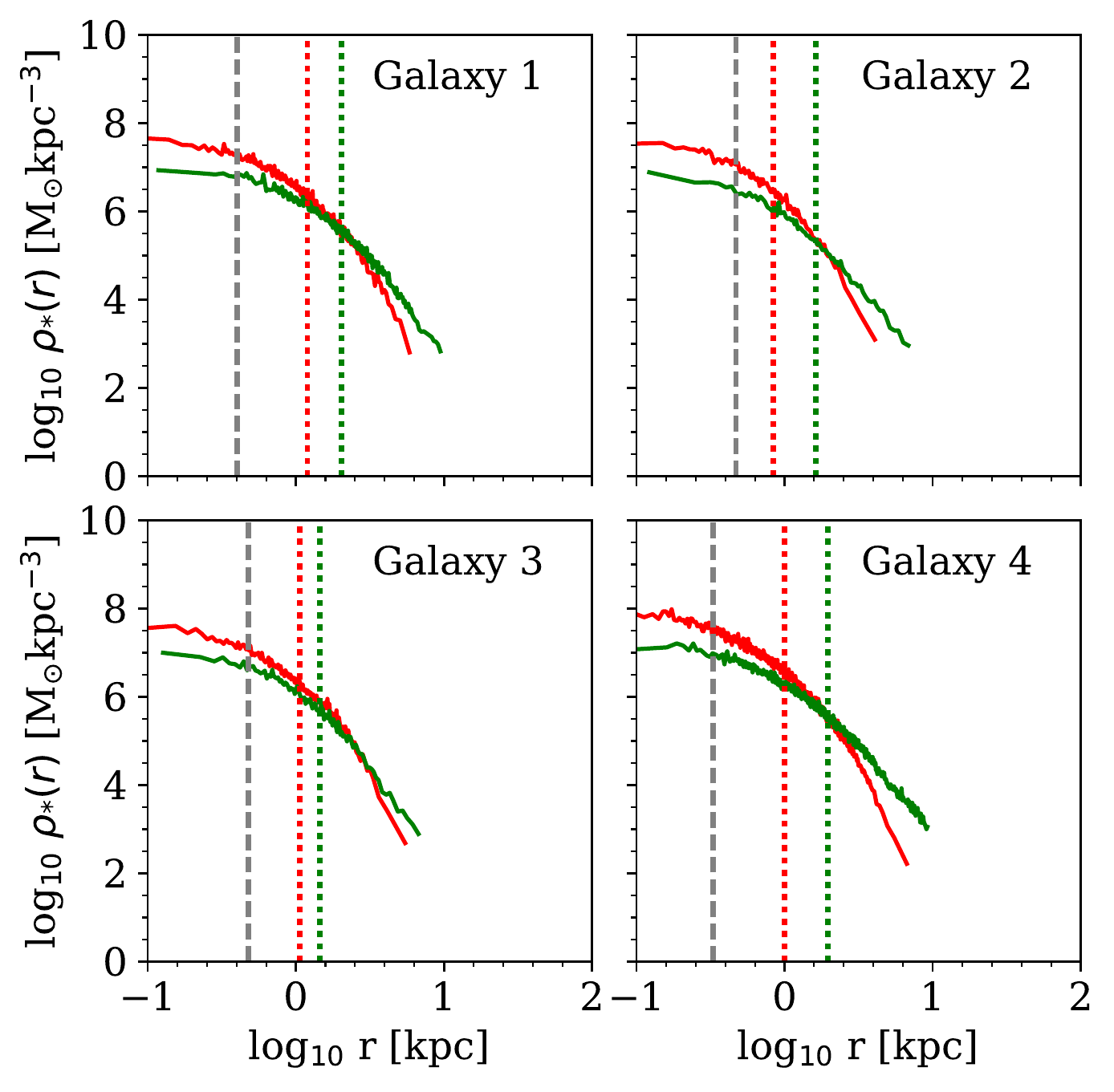}\par
    \end{multicols}
\begin{multicols}{2}
    \includegraphics[width=\linewidth]{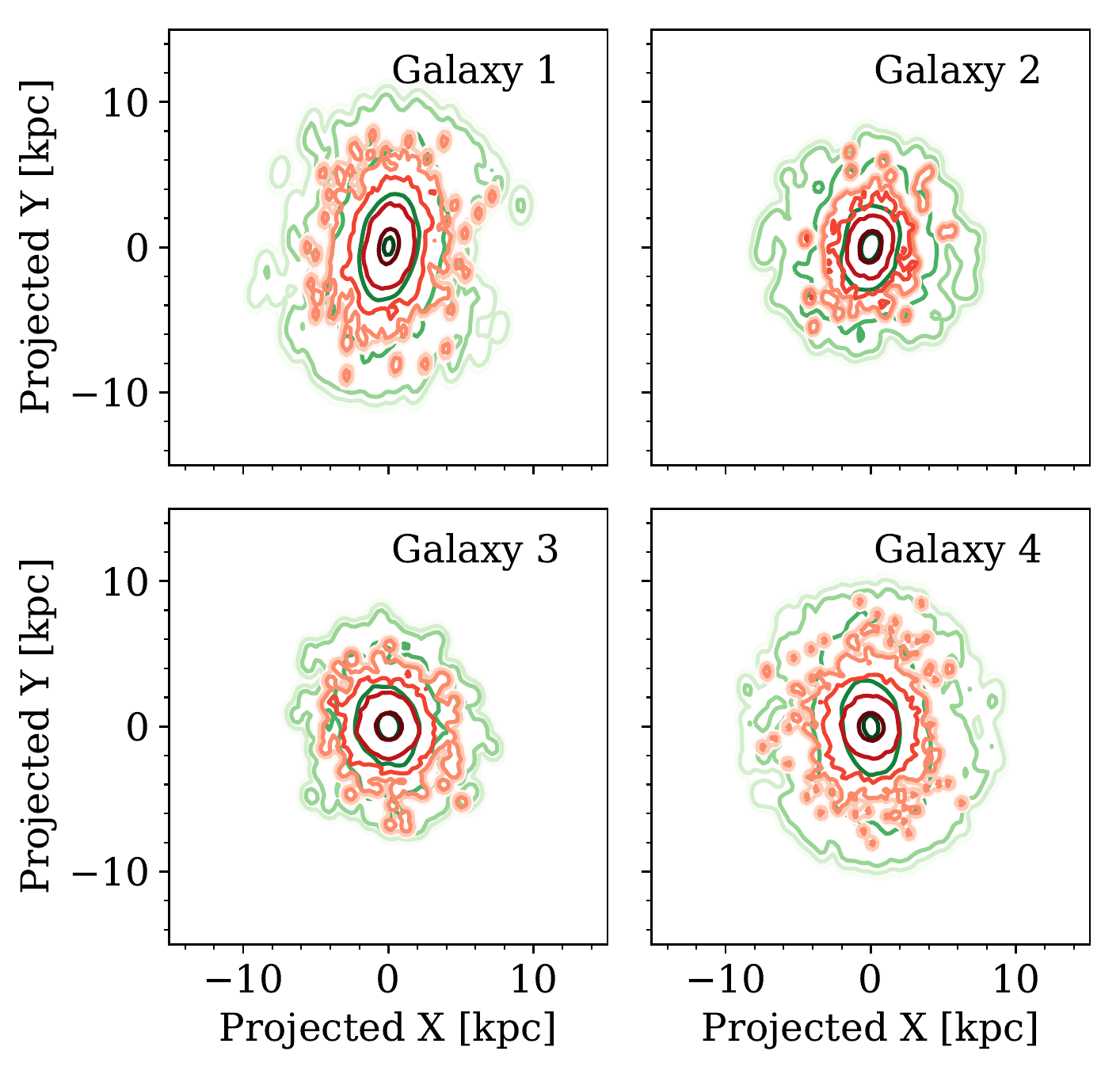}\par
    \includegraphics[width=\linewidth]{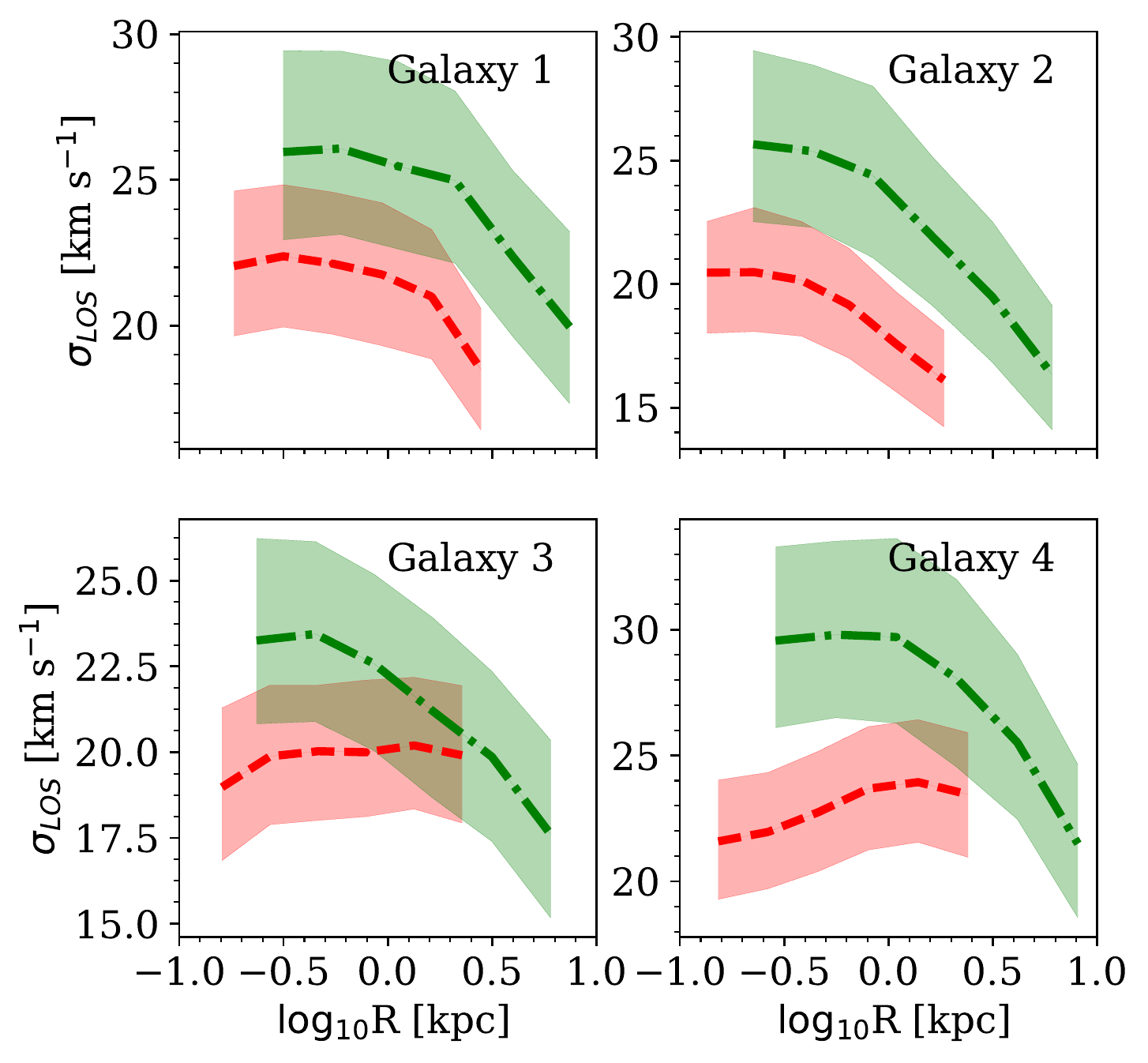}\par
\end{multicols}
\caption{\textit{Upper left:} dark matter profiles for the four
  illustrative examples discussed in this section. As for all galaxies
  in our sample, the inner density slope has a cusp; the best fitting
  NFW profile is shown in red. The grey dashed line marks the
  convergence radius defined by \citet{power}. \textit{Upper right:}
  stellar density profiles for the metal-rich (red) and metal-poor
  subpopulations (green) for the four examples. The grey dashed line
  marks the convergence radius and the red and green dotted lines show
  the 3D half-mass radii of the metal-rich and metal-poor
  subpopulations respectively.  The metal-poor component is more
  extended, with the metal-rich population concentrated near the
  centre. \textit{Lower left:}
  probability density contours of the spatial distribution of stars in
  the four galaxies. The red and green contours represent the
  metal-rich and metal-poor subpopulations, respectively, with the
  highest contour enclosing $p(x,y) = 0.1$ and lower ones decreasing
  in probability density by factors of 10. The galaxies are viewed
  along the direction of the intermediate axis of the metal-poor
  subpopulation.  \textit{Lower right:} velocity dispersion profiles
  for the four examples, projected over 100 lines of sight, as a
  function of projected distance from the centre. The red dashed line
  and the green dash-dotted line show the medians of the metal-rich
  and the metal-poor subpopulations, respectively. The bands show the
  1$\sigma$ scatter. The metal-rich subpopulation is kinematically
  colder than the metal-poor subpopulation at small radii. }
\label{char}
\end{figure*}

The four examples are highlighted in red in Fig.~\ref{LOS}.  All four
are isolated field galaxies, with no recent major mergers. Their two
metallicity subpopulations are well segregated spatially.  The
line-of-sight velocity dispersion of the metal-poor subpopulation is
higher than that of the metal-rich subpopulation (see Fig.~\ref{char}
for the properties of the four examples). In two cases (Galaxies 1 and
2) the procedure on average recovers an accurate value of the slope,
but in the other two (Galaxies 3 and 4) the procedure fails and,
instead of a cusp, it often returns a profile with a core.

\begin{figure*}

\begin{multicols}{2}
    \includegraphics[width=\linewidth]{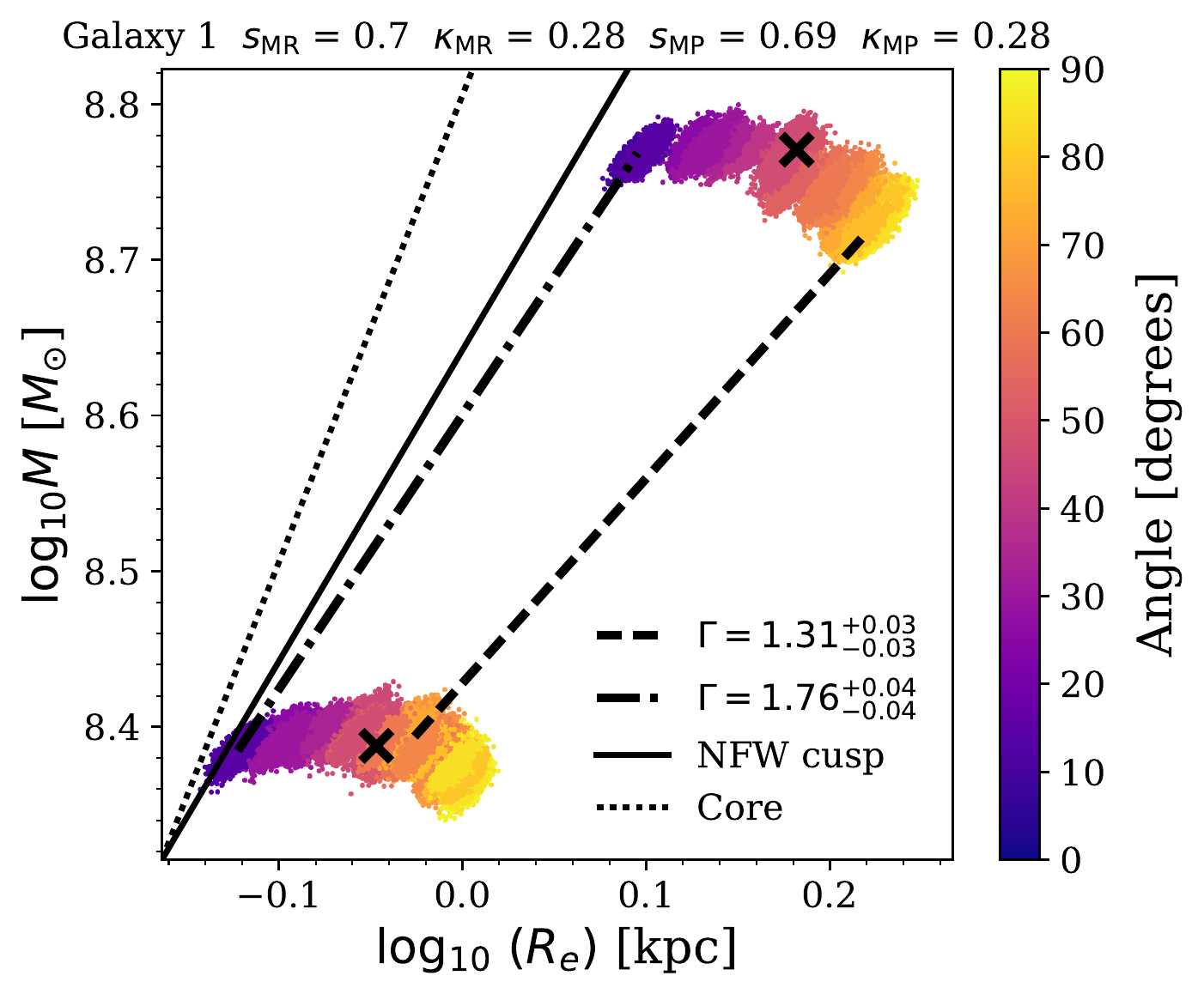}\par
    \includegraphics[width=\linewidth]{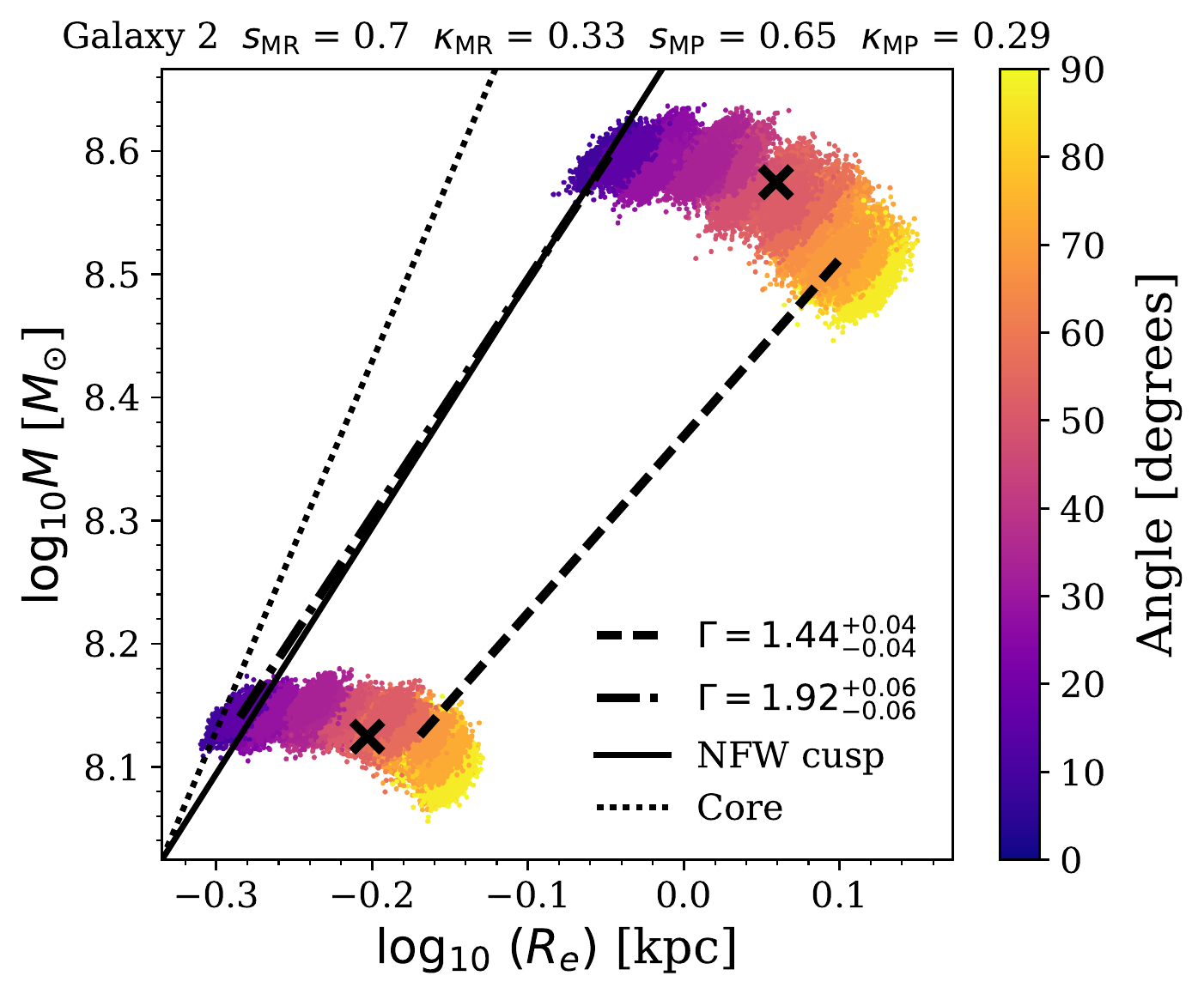}\par
    \end{multicols}
\begin{multicols}{2}
    \includegraphics[width=\linewidth]{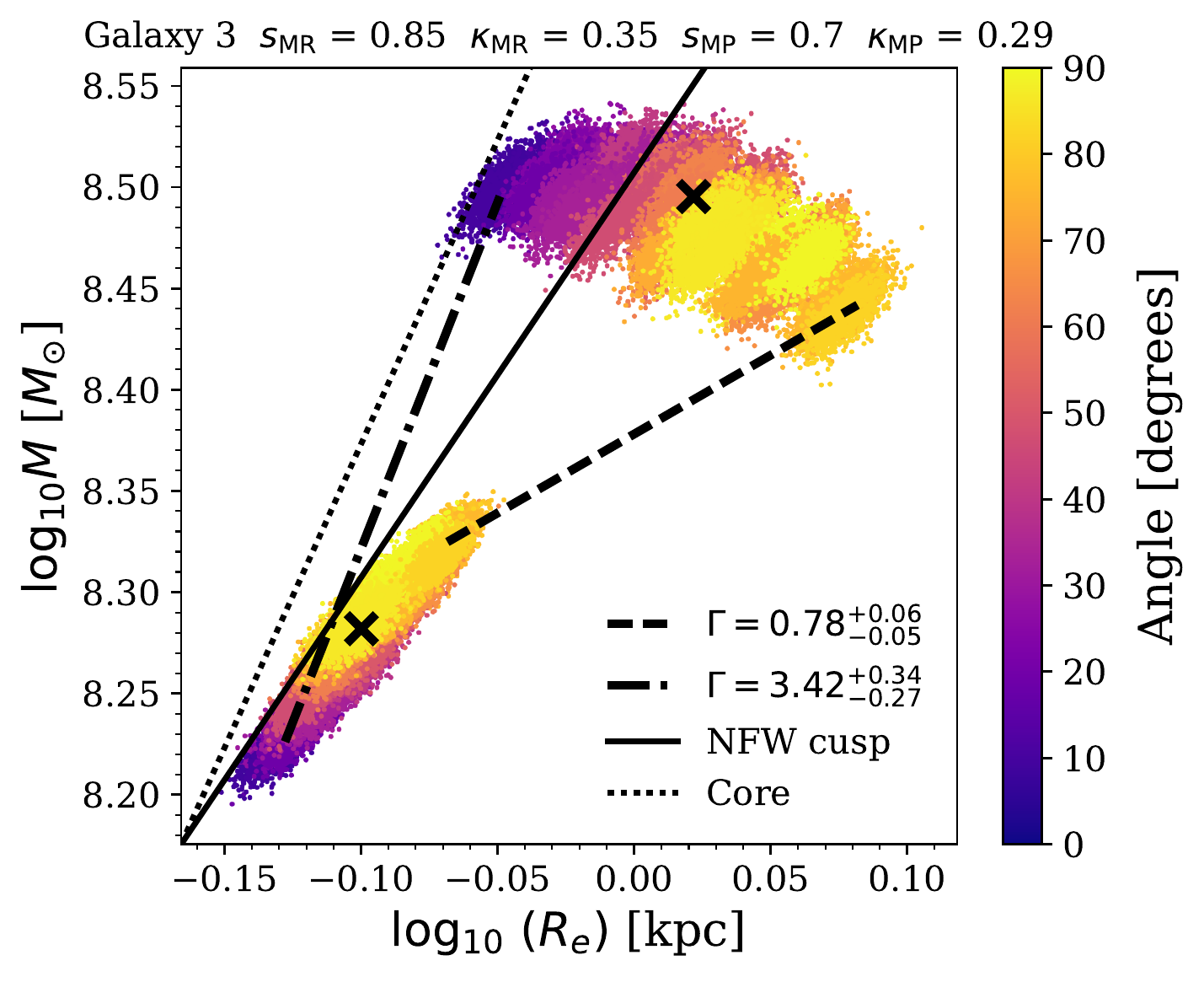}\par
    \includegraphics[width=\linewidth]{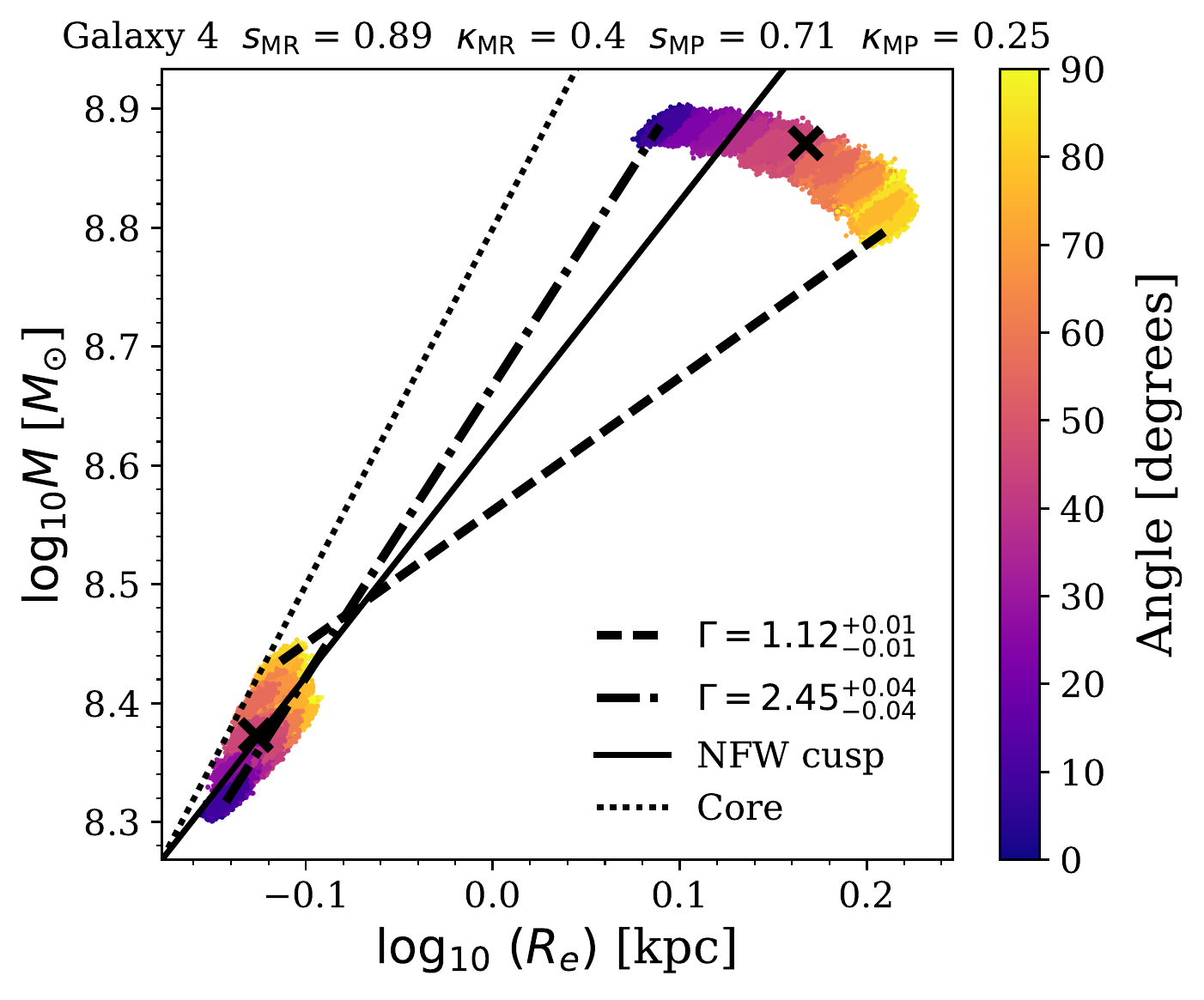}\par
\end{multicols}
\caption{Logarithmic mass slopes for our four illustrative examples.
  The dots show measured projected half-mass radii and associated contained mass inferred
  from the \citet{walkerest} estimator of equation~(\ref{walkereq}) for each of the 1000
  bootstrap resamplings of each galaxy; each galaxy is seen from 100
  different directions. The points are coloured according to the
  viewing angle measured from the major axis of the metal-poor
  subpopulation. The black dashed and dash-dotted lines show the
  minimum and maximum slopes obtained from all 100 lines of
  sight, respectively. For reference, the black solid and dotted lines
  show the slopes of an NFW cusp and a core. The black crosses denote
  the true projected 3D half-mass radii and the masses within them, taken
  directly from the simulation. The labels above each panel give the
  sphericity and value of $\kappa_{\mathrm{rot}}$ for each
  subpopulation. A large scatter in projected half-mass radius can be seen in
  subpopulations that are strongly aspherical ($s<1$). }
\label{slopesplot}
\end{figure*}

\begin{figure*}

\includegraphics[width=\linewidth]{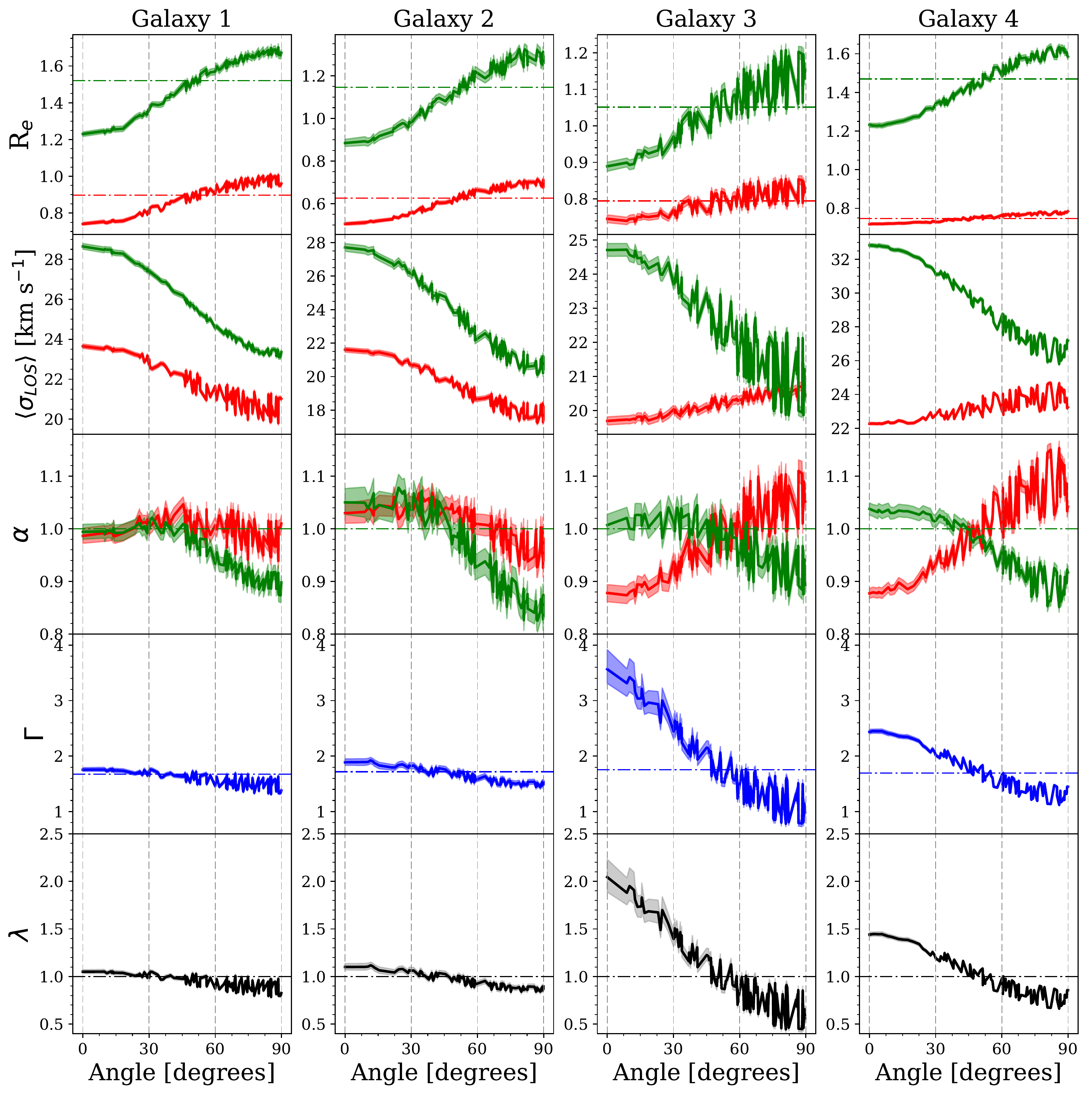}
\caption{The variation with viewing angle of some of the key
  properties of our four illustrative example galaxies.
  \textit{Angle} is the viewing angle measured from the major axis of
  the metal-poor subpopulation. Red and green colours represent the
  metal-rich and metal-poor subpopulations respectively. \textit{First
    row}: the projected half-mass radius measured for each
  subpopulation. \textit{Second row:} the estimated mass-weighted
  average line-of-sight velocity dispersion. For Galaxies 3 and 4 the velocity dispersion of the
  metal-rich subpopulation is anticorrelated with that of the
  metal-poor subpopulation. \textit{Third row:} the accuracy
  of the mass estimator for each subpopulation, $\alpha =
  M_{\mathrm{est}}/M_{\mathrm{true}}$. \textit{Fourth row:} the
  measured slope of the cumulative logarithmic mass distribution,
  $\Gamma$. \textit{Fifth row:} the accuracy of the measured mass
  slope, $\lambda = \Gamma_{\mathrm{est}}/\Gamma_{\mathrm{true}}$.}
\label{ang}

\end{figure*}

\begin{figure}

\centering
    \includegraphics[width=0.9\linewidth]{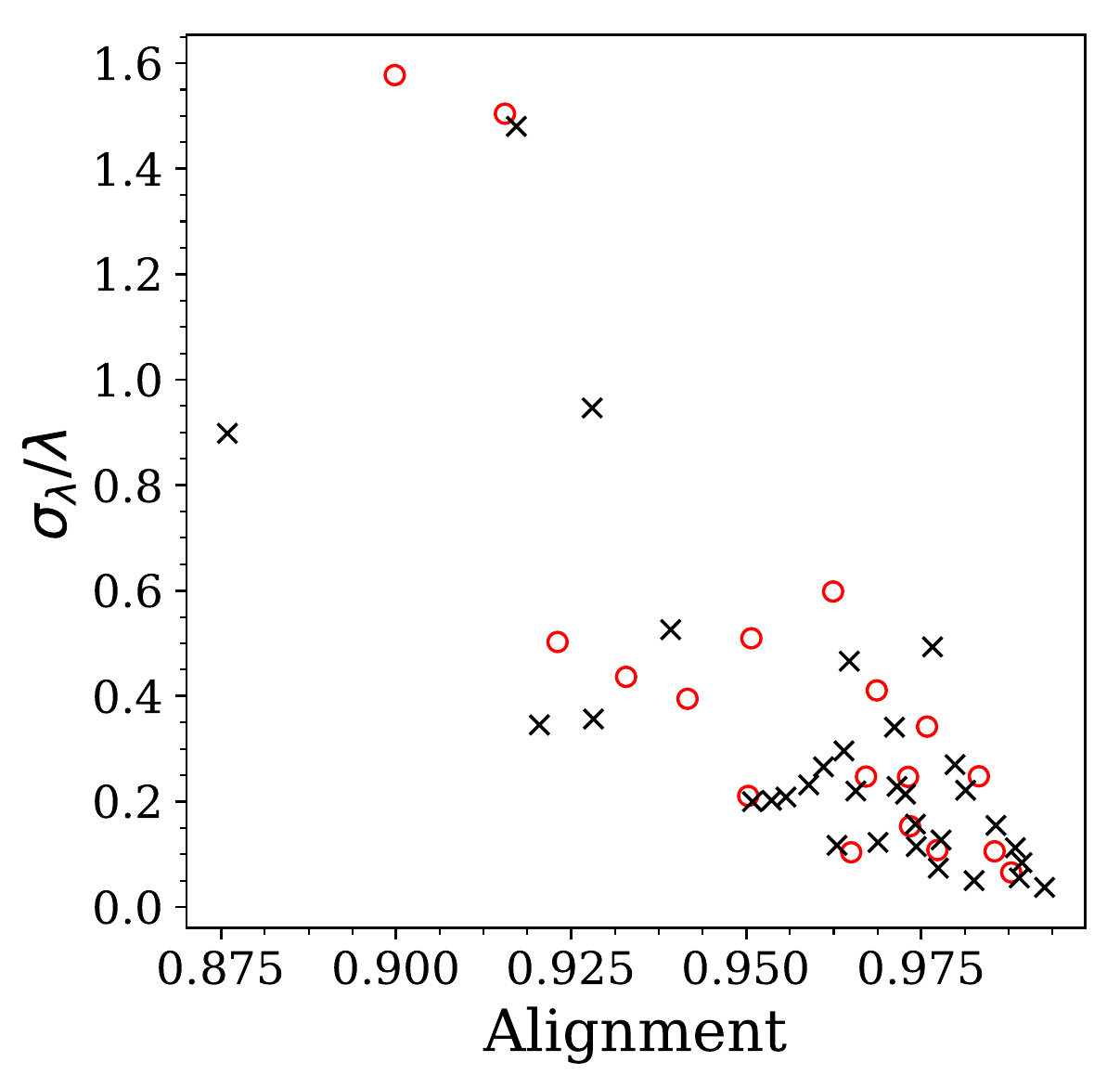}\par
    \caption{Relative upper error (the difference
        between the 84$^{\mathrm{th}}$ and the 50$^{\mathrm{th}}$
        percentile divided by the median value) on the mass slope
        accuracy, $\lambda$, as a function of the alignment of the
        metal-rich and the metal-poor subpopulations, as defined in
        Section~\ref{efsphere}. Satellites and field galaxies are
        shown with red circles and black crosses respectively.
        Measurements in galaxies with more misaligned subpopulations
        are more likely to return larger values of~~$\Gamma$.}
\label{miserr}
\end{figure}

\begin{figure}

\centering
\includegraphics[width = 0.9\linewidth]{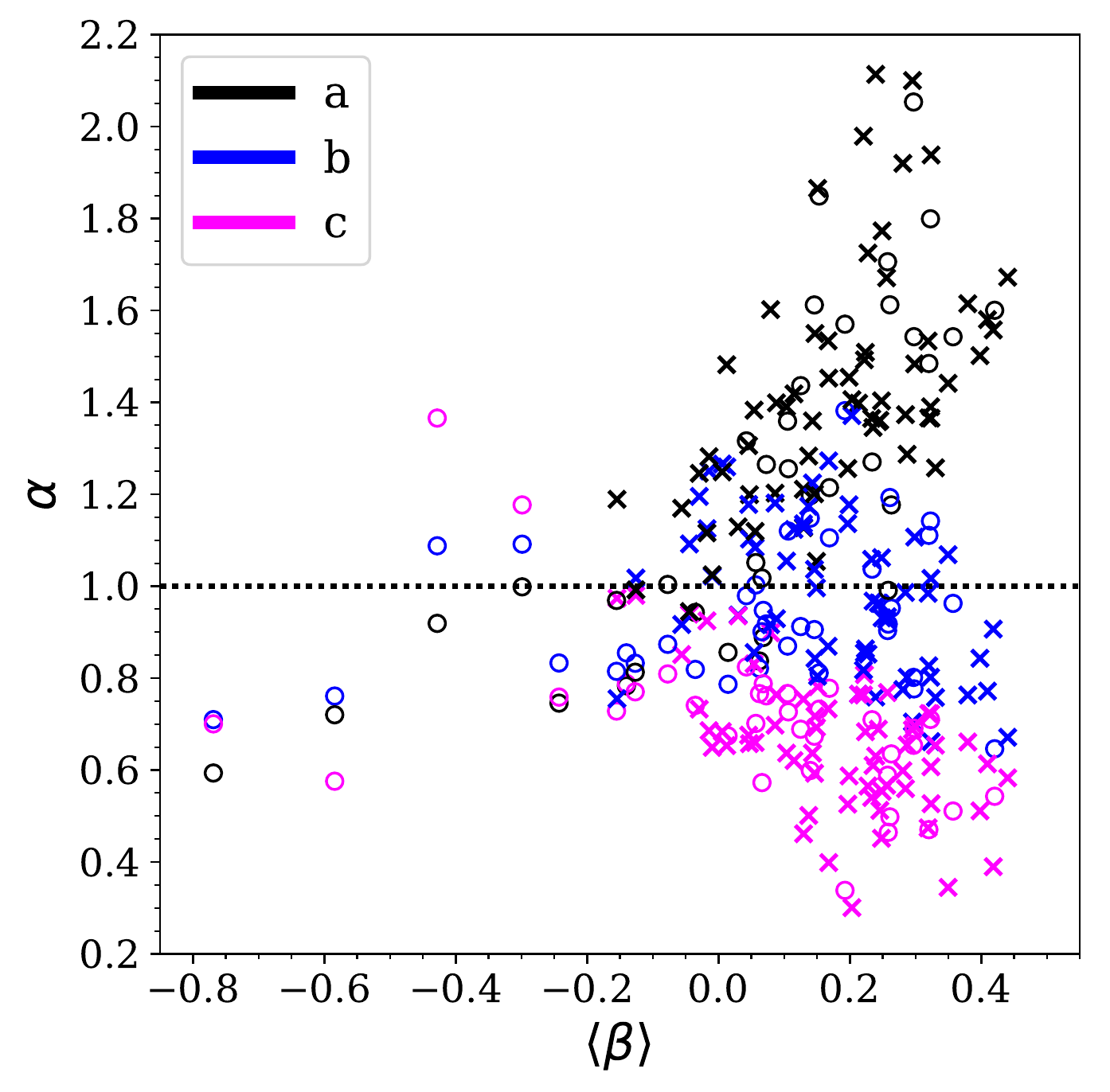} \par
\caption{The accuracy of the mass estimator, $\alpha
  =M_{\mathrm{est}}/M_{\mathrm{true}}$, for the metal-rich and
  metal-poor subpopulations of all galaxies in our sample, as a
  function of the average velocity anisotropy, $\langle \beta
  \rangle$. The galaxies are viewed along their major (black),
  intermediate (blue) and minor (magenta) axes of \textit{each}
  subpopulation. The black dotted line shows accurate mass estimates
  ($\alpha$ = 1). Satellite galaxies are shown as circles and field 
  galaxies as crosses. Along the major axis, the masses tend to be
  overestimated for subpopulations with radial anisotropy. For nearly
  isotropic subpopulations the accuracy is similar for all three
  directions. }
\label{LOSAlpha}
\end{figure}

\begin{figure*}
\centering
  \includegraphics[width=2.\columnwidth]{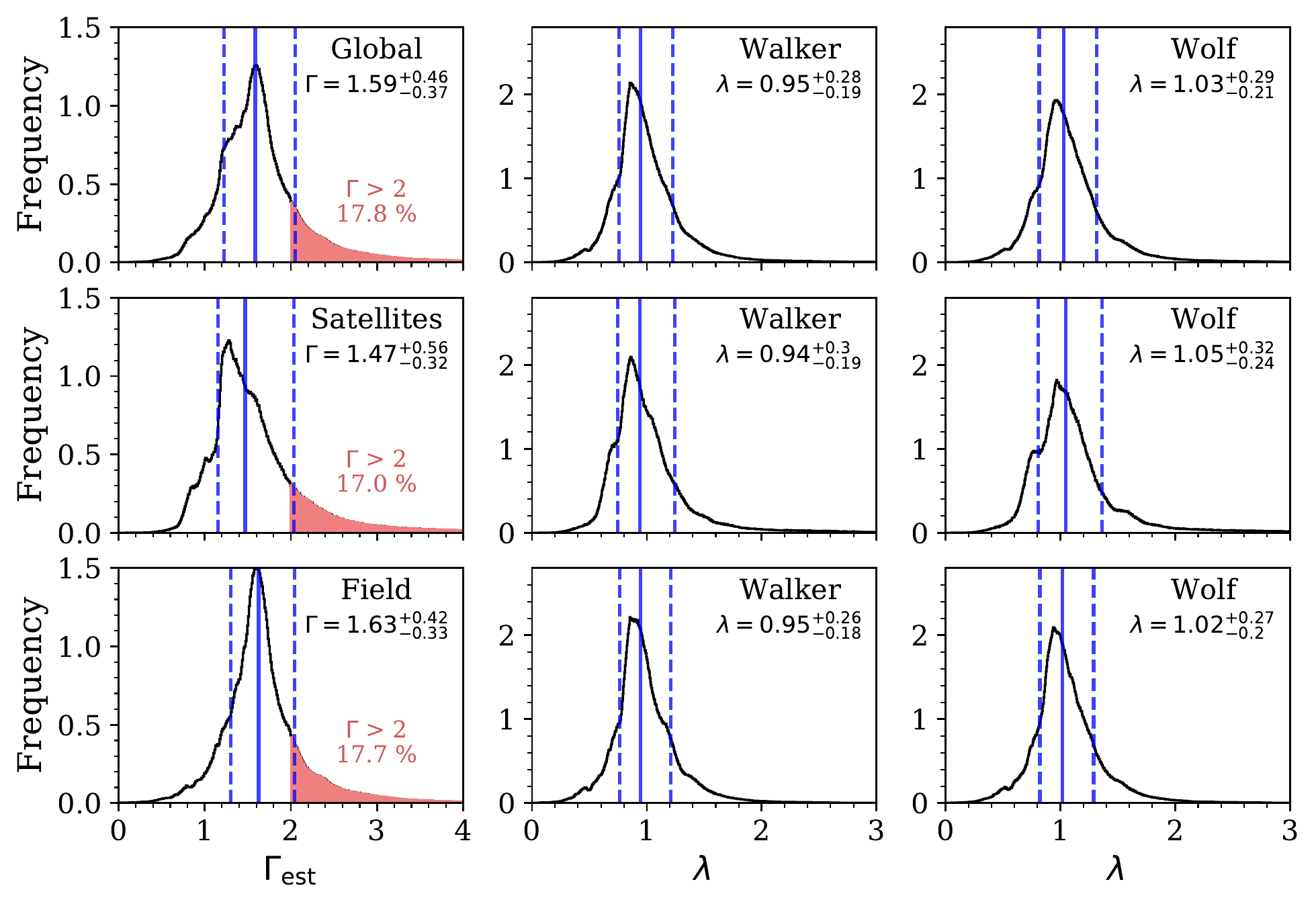}\par

  \caption{\textit{Top panel:} overall distributions for the 50 objects in our sample of the measured slope, $\Gamma_{\mathrm{est}}$ (left), and slope
      accuracy, $\lambda=
      \Gamma_{\mathrm{est}}/\Gamma_{\mathrm{true}}$, for (middle) the
      \citet{walkerest} estimator and (right) the \citet{wolf} estimator.
       Blue solid lines show the
      median values and the blue dashed lines show 16$^{\mathrm{th}}$
      and 84$^{\mathrm{th}}$ percentiles. The slopes tend to be
      underestimated by 5 per cent on average using the
      \citet{walkerest} estimator and overestimated by 3 per cent on
      average using the \citet{wolf} estimator; the distribution of
      $\lambda$ is not symmetrical. $\Gamma \geq 2$ is measured in
      17.8 per cent of cases and $\Gamma \geq 3$ in $\sim$ 3.5 per
      cent of cases. \textit{Middle panel:} same as above, but
      only for the 18 satellite galaxies in our
      sample. \textit{Bottom panel:} as above, but showing the
      distributions of $\Gamma$ and $\lambda$ only for the 32 field
      galaxies in our sample. The estimators perform in a similar way
      for both satellites and field galaxies.}
 \label{global}
\end{figure*}

\subsection{The effects of misalignment and anisotropy}
\label{efsphere}
We view each galaxy from 100 random directions generated uniformly on
the surface of a sphere using the spherical spiral method outlined in
\citet{fibonacci}. As in Section~\ref{sec:WP}, for each line of sight
we generate 1000 bootstrap resamplings (with replacement) of the stars
in each galaxy and obtain median values of the quantities of
interest. In Fig.~\ref{slopesplot}, which is analogous to fig.~10 of
\cite{walk}, we plot the logarithm of the measured projected half-mass radii $R_{\mathrm{e}}$ of the bootstrap
samples for each viewing angle, along with the corresponding logarithm of masses
contained within these radii, inferred using the estimator of
equation~(\ref{walkereq}). The measurements are coloured according to the
angle between the line of sight and the major axis of the
\textit{metal-poor} subpopulation. (The major axes of the two
subpopulations will not necessarily be aligned.) The true projected 3D half-mass
radius and the enclosed mass of each subpopulation are marked by a
cross. The sphericity and value of $\kappa_{\mathrm{rot}}$ for each
metallicity subpopulation are given in the legend of each panel. For
each viewing angle the inferred slope of the cumulative logarithmic
mass profile can be obtained by joining points of the same colour in
the metal-rich and metal-poor subpopulations. The dashed and
dot-dashed lines show the minimum and maximum slopes inferred by this
procedure.

The variation with viewing angle of relevant quantities
  for each galaxy is plotted in Fig.~\ref{ang} and is discussed
  below. In this plot, $\alpha$ denotes the ratio,
  $M_{\mathrm{est}}/M_{\mathrm{true}}$, of the estimated to the true
  mass, where the true mass includes the dark matter, stellar and gas
  particles within the projected 3D half-mass radius; $\lambda$ is the
  ratio, $\Gamma_{\mathrm{est}}/\Gamma_{\mathrm{true}}$, of the
  estimated to the true slope of the cumulative logarithmic mass
  profile, where $\Gamma_{\mathrm{est}}$ is the slope measured between
  the measured values of $R_{\mathrm{e}}$ and $\Gamma_{\mathrm{true}}$ is the
  slope between the projected true 3D half-mass radii of the two
  subpopulations.

Apart from the metal-rich subpopulation of Galaxy~4, which is close to
spherical, all other subpopulations in the four examples of
Fig.~\ref{slopesplot} are quite elongated. For the nearly spherical
subpopulation, the estimated projected half-mass radius is, not surprisingly,
almost independent of viewing angle. The values of the estimated mass
show a mild systematic dependence on viewing angle which reflects the
weak dependence of the velocity dispersion on the line of sight seen
in Fig.~\ref{ang}. 

The situation is quite different for the elongated
  subpopulations. Let us consider, for example, the metal-poor
  subpopulation of Galaxy~4. Its sphericity is $s_{\rm{MP}}=0.71$,
  quite typical for our sample (see Fig.~\ref{properties}). Now the
  measured values of the projected half-mass radius vary greatly with
  viewing angle, from 1.2~kpc when the subpopulation is viewed along
  its major axis to 1.6~kpc when it is viewed along the perpendicular
  direction. Since the velocity anisotropy is radially biased (see
  Fig.~\ref{properties}), the measured velocity dispersion varies with
  viewing angle, from 33~km~s$^{-1}$ when viewed along the major axis
  to 27~km~s$^{-1}$ when viewed along the perpendicular
  direction. This decrease in velocity dispersion largely balances the
  increase in the projected half-mass radius with the result that the
  estimated mass of the metal-poor population varies little with
  viewing angle, by only $\sim$~0.1 dex. Thus, a very similar mass is
  associated with a relatively large range of projected half-mass
  radii. The biases in the measured mass are anticorrelated as the
  viewing angle varies, as shown in the third row of Fig.~\ref{ang}.

As Figs~\ref{slopesplot} and~\ref{ang} show, the result is that the
inferred slope of the cumulative logarithmic mass profile for Galaxy~4
can take on a wide range of values, from $\Gamma=2.45$ when the galaxy is
viewed along the major axis of the metal-poor population to $\Gamma=1.12$ when it is viewed along the
perpendicular direction. Thus, for viewing angles between about 0 and 50$^{\circ}$, the mass distribution in this galaxy would, incorrectly, be
measured to have a core.

Galaxies 1 and 2 are also quite elongated. However, in these two
cases, both subpopulations are aspherical and their major axes are
roughly aligned. Now the behaviour we have just seen for the
metal-poor subpopulation of Galaxy~4 is reproduced {\em for both
  subpopulations}. The result is that both the projected half-mass radius and
the mass contained within it for both subpopulations are incorrectly
estimated by roughly the same factors and these biases vary similarly
with viewing angle (third row of Fig.~\ref{ang}).  Thus, although the
masses and radii of both subpopulations are incorrectly estimated, the
slopes come out roughly right: these two galaxies are correctly
inferred to have cusps.

Galaxy~3 is an intermediate case. Its metal-poor subpopulation has
similar sphericity to the metal-poor subpopulation of Galaxy~4 so the
measured projected half-mass radius varies by a similar factor. While the inferred
mass has greater scatter at a given viewing angle, the overall
variation is still only slightly larger than 0.1 dex. The
half-mass radius of the metal-rich subpopulation of Galaxy~3 shows
some dependence on viewing angle (see Fig.~\ref{ang}). In this case,
the systematic variation of the inferred mass is larger and, as seen
in Fig.~\ref{ang}, it increases with viewing angle. This
is enough to result in a wide range of estimated slopes. Similarly to
Galaxy~4, this galaxy would be measured to have a core for viewing
angles between about 20 and 50$^{\circ}$ and for even smaller viewing
angles it would be measured to have a `hole' in the central region
($\Gamma > 3$). 

Although the systematic errors in the estimates of the slope of the
cumulative logarithmic mass distribution using a procedure analogous
to that of \citet{walk} result from a complex interplay between
projection and kinematic effects, it is clear that a major factor
behind them is the significant elongation of the stellar
subpopulations, and particularly the misalignment between the major
axes of the metal-rich and metal-poor subpopulations seen in a number
of cases, including Galaxy~4. This lack of similarity arises naturally
in our simulations and is linked to the different formation paths of
the two subpopulations \citep{mergers}, which we will investigate in a
subsequent paper. 

\pagebreak

In order to quantify the effect of both the
  misalignment of the principal axes and the differences in sphericity
  of the two subpopulations on the scatter in the estimated
  logarithmic mass slopes, we define the following alignment
  statistic. We model the two subpopulations as concentric ellipsoids
  of unit volume, with axis ratios equal to those measured for each
  subpopulation and an offset angle equal the angle between the major
  axes of the two subpopulations. 
  We then compute the fraction of volume within the intersection of
  the two ellipsoids. Subpopulations with similar axis ratios and
  spatial orientation would therefore have an alignment close to 1. In
  Fig.~\ref{miserr} we show the relative upper error (the difference
  between the 84$^{\mathrm{th}}$ and 50$^{\mathrm{th}}$ percentiles
  divided by the median value)
  on the slope accuracy, $\lambda$, as a function of the value of the
  alignment statistic. It is clear that measurements for galaxies with
  more misaligned subpopulations tend to return higher values of
  $\Gamma$, corresponding to shallower inner density slopes.

The non-trivial radial variation of the velocity
anistropy, $\beta$, which is generally different for the two
metallicity subpopulations (see Fig.~\ref{properties}), also plays a
role. This can be seen in Fig.~\ref{LOSAlpha}, where we plot $\alpha$, the
error in the estimate of the mass for each of the two subpopulations in our
sample, as a function of the average anisotropy parameter $\langle
\beta \rangle$, when each subpopulation is viewed from directions
aligned with the three principal axes. For directions along the minor
and intermediate axes the values of $\alpha$ scatter about $\alpha=1$,
although there is a bias towards $\alpha<1$, that is for an
underestimate of the mass when the subpopulation is viewed along its
minor axis. However, when the subpopulation is viewed along its major
axis and the velocity anisotropy has a radial bias ($\beta>0$) , its
mass tends to be {\em overestimated} [in agreement with the
conclusions of \citet{orientation}] and the size of this bias increases
systematically with increasing anisotropy. Along this particular
viewing angle the velocity dispersion is generally largest.  When the
velocity distribution is isotropic or tangentially biased, however,
the mass is correctly estimated. (This is the reason why the estimate
of the mass of the metal-rich subpopulation of Galaxy 3, which has an
isotropic velocity distribution, is unbiased.)

\subsection{Accuracy of the inferred mass slope for the sample as a whole}
\label{general}

In this subsection we present statistical results for our sample of 50
galaxies. This sample was selected according to the specific criteria
described in Section~\ref{splitting}, essentially requiring that there
be two well-separated metallicity subpopulations, as judged by the
statistic introduced in that section. These criteria need not match in
detail those used for real objects. At present there are only a
handful of two-metallicity subpopulation galaxies known and those
populations have been identified somewhat
serendipitously. Nevertheless, our statistical results should be
indicative of the frequency with which we might expect the slope of
the dark matter density profile to be incorrectly estimated.

The top panel of Fig.~\ref{global} shows the
  distributions of $\Gamma_{\mathrm{est}}$ and
  $\lambda=\Gamma_{\mathrm{est}}/\Gamma_{\mathrm{true}}$ for the 1000
  bootstrap resamplings of each galaxy in our sample, each viewed from
  100 different random directions. (Recall that an NFW profile has
  $\Gamma=2$ as $r \xrightarrow {}0$ and a profile with a constant-density core has $\Gamma=3$.) The distribution of
  $\Gamma_{\mathrm{est}}$ is asymmetric with mean value $\Gamma$~=~
  1.59$^{+0.46}_{-0.37}$ and $\lambda$ = 0.95$^{+0.28}_{-0.19}$ for
  the \citet{walkerest} estimator, implying that $\Gamma$ is
  underestimated on average by $\sim$ 5 per cent; the true value
  (corresponding to $\lambda=1$) lies within 1$\sigma$ of the
  mean. Slopes tend to be overestimated by $\sim$ 3 per cent on
  average using the \citet{wolf} estimator, with $\lambda$ =
  1.03$^{+0.29}_{-0.21}$. 

Flatter slopes than NFW, $\Gamma >2$ are measured in
  $\sim 17.8$ per cent of cases and $\Gamma > 3$ (corresponding to a
  `hole' in the centre) in $\sim 3.5$ per cent of cases. In the middle
  and bottom panels we show equivalent distributions for satellite and
  field galaxies in our sample. Whilst the satellites generally tend
  to exhibit cuspier inner slopes
  ($\Gamma_{\mathrm{est}}$~=~1.47$^{+0.56}_{-0.32}$ compared to
  $\Gamma_{\mathrm{est}}$~=~1.63$^{+0.42}_{-0.33}$ for field dwarfs),
  the distribution of the accuracy of the inferred slopes, $\lambda$,
  is similar to that of the field galaxies. The greater scatter
  towards higher values of $\Gamma$ in satellites can be explained by
  the fact that a greater fraction of satellites than field galaxies
  in our sample exhibit strong misalignment, as shown in
  Fig.~\ref{miserr}. We have additionally verified that if we identify
  the true slope with the slope of the line joining the logarithm of true masses within $R_e$ of each subpopulation, the bias and scatter in $\lambda$ remain unaffected.

\citet{walk}, who used the \citet{walkerest} estimator, found $\Gamma=2.95^{+0.51}_{-0.39} $ for Sculptor and
  $\Gamma=2.61^{+0.43}_{-0.37} $ for Fornax and concluded that these
  values exclude the NFW profile with significance greater than 99 and
  96 per cent respectively. However, according to the distribution of
  $\Gamma_{\rm est}$ in Fig.~\ref{global} for the \citet{walkerest}
  and \citet{wolf} estimators derived from our simulations, these
  values of $\Gamma$ are only inconsistent with the NFW profile at
  93.6 and 88.9 significance for Sculptor and Fornax respectively
  (where we have taken the 1$\sigma$ lower limit of the original
  estimates).
\section{Discussion and Conclusions}
\label{conc}
The question of whether or not the dark matter haloes of galaxies have
central, constant-density cores is of great interest in cosmology. It
is now well established that, in the absence of baryon effects, haloes
of all masses develop NFW profiles \citep{navarro1996b,nfw} which have
a central cusp. An incontrovertible measurement of a core would
therefore have important implications: it would either signal the
impact of exotic baryonic effects \citep{navarro1996b,pontzen} or of
exotic types of particles such as self-interacting dark matter
\citep{selfintdm}. Unfortunately measuring dark matter profiles in the
innermost regions of haloes, encompassing only at most a few per cent
of the halo mass, is difficult.

In this work we have tested the \citet{walk} procedure for inferring
the slope of the inner dark matter halo profile in galaxies with two
metallicity subpopulations using galaxies from the APOSTLE
cosmological simulations. These follow not only the evolution of dark
matter, but also the evolution of gas, and incorporate subgrid
prescriptions to model the processes thought to be at play in galaxy
formation. The initial conditions correspond to a cold dark matter
universe but are conditioned to form an analogue of the Local Group at
the final time. Thus, this is the first test of the \citet{walk}
procedure in fully realistic simulations which produce dwarf galaxies
{\em ab initio}. These are not guaranteed to satisfy the key
assumptions underlying the \citet{wolf} and \citet{walkerest} mass
estimators: sphericity and dynamical equilibrium, and indeed, many of
our simulated galaxies violate them to varying degrees.

Gratifyingly, our simulations produce dwarf galaxies with two
identifiable metallicity subpopulations reminiscent of Sculptor and
Fornax. Out of 286 model galaxies resolved with more than 1000 stellar
particles we identify 50 dwarfs with dual metallicity subpopulations,
according to the criteria discussed in Section \ref{splitting}.  Their
kinematical properties are summarized in Fig.~\ref{properties}.  The
subpopulations tend to be significantly aspherical and often develop
different asphericities: $\sim 30$ per cent of galaxies in our sample
have metal-rich and metal-poor subpopulations differing in sphericity
by over 10~per cent and, in many cases, the orientation of their major
axes differs as well.  The metal-rich subpopulations tend to show more
rotation than the metal-poor ones and the velocity anisotropy of the
subpopulations can also differ substantially. These differences occur
both in satellites and in field dwarfs. The lack of similarity arises
naturally in our simulations and is linked to the different formation
paths of the two subpopulations (which we will investigate in a
subsequent paper).

Our main result is that the method introduced by \cite{battaglia} and
extended by \cite{walk}, based on comparing the masses interior
to the half-mass radii of the metal-rich and metal-poor
subpopulations, can often lead to an incorrect inference of the slope
of the inner profile of the galaxies' dark matter haloes. All the haloes
in our simulations have cuspy NFW profiles, yet 17.8~per~cent of the galaxies in
our sample would be inferred to have flatter, core-like
profiles. Multiple factors play a role in these failures but the main
culprit is misalignment of the two elongated metallicity
subpopulations (which generally have radially varying velocity
anisotropy). This results in a wide range of inferred slopes for the
mass distribution depending on the viewing angle. Four specific
examples illustrating cases when the slope is correctly or incorrectly
inferred have been shown.

The study of \citet{walk} is sometimes regarded as one of the more
convincing arguments for the presence of cores in the dark matter
haloes of dwarf galaxies \citep{amorisco}. However, their proposed
method of measuring cumulative logarithmic mass profile slopes relies
heavily on the assumption that the properties of the two metallicity
subpopulations are correlated, such that any bias in the application
of the mass estimator would cancel when obtaining the slope. This
assumption fails in a number of our simulated galaxies where not only
the elongation and orientation, but also the orbital anisotropy
differs for the two subpopulations.  Of course it remains to be seen
whether the properties of the two metallicity subpopulations in
Sculptor and Fornax are correlated or not, but this is not currently
feasible. The availability of proper motion data for individual stars within these galaxies would allow us to somewhat take away the dependence of the accuracy of the mass estimator on the angle of view, although, for an instrument like Gaia, the measurement uncertainties for a galaxy as distant as Sculptor is expected to be comparable to the velocities themselves \citep{gaia,sculptorproper}. Given the sensitivity of the inferred slope to the viewing
angle that we have demonstrated in this work, we concur with
\citet{orientation} that a large statistical sample of
\textit{randomly oriented} galaxies is highly desirable. But for any
particular observation, knowledge of the sphericity and velocity
anisotropy of both subpopulations and their orientation with respect
to one another and to the observer is required before any definitive
conclusion can be reached regarding the slope of the galaxy's inner
dark matter halo.

\section*{Acknowledgments}
We thank the anonymous referee for useful comments that helped us improve
the final version of this work.  We are also grateful for helpful
discussions with Marius Cautun, Aaron Ludlow, Matthieu Schaller, Louis
Strigari and Alis Deason. This work was supported by the Science and
Technology Facilities Council grants ST/L00075X/1 and ST/P000451/1. AG acknowledges an STFC
studentship funded by STFC grant ST/N50404X/1. This work used the
DiRAC Data Centric system at Durham University, operated by the
Institute for Computational Cosmology on behalf of the STFC DiRAC HPC
Facility (www.dirac.ac.uk). This equipment was funded by BIS National
E-infrastructure capital grant ST/K00042X/1, STFC capital grant
ST/H008519/1, STFC DiRAC Operations grant
ST/K003267/1 and Durham University. DiRAC is part of the National
E-Infrastructure. This work has also benefited from the use of {\sc{numpy}}
and {\sc{scipy}}.




\bibliographystyle{mnras}
\bibliography{metal} 

\begin{thebibliography}{}
\makeatletter
\relax
\def\mn@urlcharsother{\let\do\@makeother \do\$\do\&\do\#\do\^\do\_\do\%\do\~}
\def\mn@doi{\begingroup\mn@urlcharsother \@ifnextchar [ {\mn@doi@}
  {\mn@doi@[]}}
\def\mn@doi@[#1]#2{\def\@tempa{#1}\ifx\@tempa\@empty \href
  {http://dx.doi.org/#2} {doi:#2}\else \href {http://dx.doi.org/#2} {#1}\fi
  \endgroup}
\def\mn@eprint#1#2{\mn@eprint@#1:#2::\@nil}
\def\mn@eprint@arXiv#1{\href {http://arxiv.org/abs/#1} {{\tt arXiv:#1}}}
\def\mn@eprint@dblp#1{\href {http://dblp.uni-trier.de/rec/bibtex/#1.xml}
  {dblp:#1}}
\def\mn@eprint@#1:#2:#3:#4\@nil{\def\@tempa {#1}\def\@tempb {#2}\def\@tempc
  {#3}\ifx \@tempc \@empty \let \@tempc \@tempb \let \@tempb \@tempa \fi \ifx
  \@tempb \@empty \def\@tempb {arXiv}\fi \@ifundefined
  {mn@eprint@\@tempb}{\@tempb:\@tempc}{\expandafter \expandafter \csname
  mn@eprint@\@tempb\endcsname \expandafter{\@tempc}}}

\bibitem[\protect\citeauthoryear{{Adams} et~al.,}{{Adams}
  et~al.}{2014}]{adamssimon}
{Adams} J.~J.,  et~al., 2014, \mn@doi [\apj] {10.1088/0004-637X/789/1/63},
  \href {http://adsabs.harvard.edu/abs/2014ApJ...789...63A} {789, 63}

\bibitem[\protect\citeauthoryear{{Agnello} \& {Evans}}{{Agnello} \&
  {Evans}}{2012}]{agnello}
{Agnello} A.,  {Evans} N.~W.,  2012, \mn@doi [\apjl]
  {10.1088/2041-8205/754/2/L39}, \href
  {http://adsabs.harvard.edu/abs/2012ApJ...754L..39A} {754, L39}

\bibitem[\protect\citeauthoryear{Akaike}{Akaike}{1998}]{Akaike}
Akaike H.,  1998, Information Theory and an Extension of the Maximum Likelihood
  Principle.
Springer New York, New York, NY, pp 199--213,
  \mn@doi{10.1007/978-1-4612-1694-0_15}

\bibitem[\protect\citeauthoryear{{Amorisco} \& {Evans}}{{Amorisco} \&
  {Evans}}{2012}]{amorisco}
{Amorisco} N.~C.,  {Evans} N.~W.,  2012, \mn@doi [\mnras]
  {10.1111/j.1365-2966.2011.19684.x}, \href
  {http://adsabs.harvard.edu/abs/2012MNRAS.419..184A} {419, 184}

\bibitem[\protect\citeauthoryear{{Barber}, {Starkenburg}, {Navarro}  \&
  {McConnachie}}{{Barber} et~al.}{2015}]{barber}
{Barber} C.,  {Starkenburg} E.,  {Navarro} J.~F.,   {McConnachie} A.~W.,  2015,
  \mn@doi [\mnras] {10.1093/mnras/stu2494}, \href
  {http://adsabs.harvard.edu/abs/2015MNRAS.447.1112B} {447, 1112}

\bibitem[\protect\citeauthoryear{{Battaglia} et~al.,}{{Battaglia}
  et~al.}{2006}]{battaglia06}
{Battaglia} G.,  et~al., 2006, \mn@doi [\aap] {10.1051/0004-6361:20065720},
  \href {http://adsabs.harvard.edu/abs/2006A%26A...459..423B} {459, 423}

\bibitem[\protect\citeauthoryear{{Battaglia}, {Helmi}, {Tolstoy}, {Irwin},
  {Hill}  \& {Jablonka}}{{Battaglia} et~al.}{2008}]{battaglia}
{Battaglia} G.,  {Helmi} A.,  {Tolstoy} E.,  {Irwin} M.,  {Hill} V.,
  {Jablonka} P.,  2008, \mn@doi [\apjl] {10.1086/590179}, \href
  {http://adsabs.harvard.edu/abs/2008ApJ...681L..13B} {681, L13}

\bibitem[\protect\citeauthoryear{{Battaglia}, {Tolstoy}, {Helmi}, {Irwin},
  {Parisi}, {Hill}  \& {Jablonka}}{{Battaglia} et~al.}{2011}]{battagliasextans}
{Battaglia} G.,  {Tolstoy} E.,  {Helmi} A.,  {Irwin} M.,  {Parisi} P.,  {Hill}
  V.,   {Jablonka} P.,  2011, \mn@doi [\mnras]
  {10.1111/j.1365-2966.2010.17745.x}, \href
  {http://adsabs.harvard.edu/abs/2011MNRAS.411.1013B} {411, 1013}

\bibitem[\protect\citeauthoryear{{Ben{\'{\i}}tez-Llambay}, {Navarro}, {Abadi},
  {Gottl{\"o}ber}, {Yepes}, {Hoffman}  \& {Steinmetz}}{{Ben{\'{\i}}tez-Llambay}
  et~al.}{2016}]{mergers}
{Ben{\'{\i}}tez-Llambay} A.,  {Navarro} J.~F.,  {Abadi} M.~G.,  {Gottl{\"o}ber}
  S.,  {Yepes} G.,  {Hoffman} Y.,   {Steinmetz} M.,  2016, \mn@doi [\mnras]
  {10.1093/mnras/stv2722}, \href
  {http://adsabs.harvard.edu/abs/2016MNRAS.456.1185B} {456, 1185}

\bibitem[\protect\citeauthoryear{{Bett}}{{Bett}}{2012}]{sphericity}
{Bett} P.,  2012, \mn@doi [\mnras] {10.1111/j.1365-2966.2011.20258.x}, \href
  {http://adsabs.harvard.edu/abs/2012MNRAS.420.3303B} {420, 3303}

\bibitem[\protect\citeauthoryear{{Binney} \& {Mamon}}{{Binney} \&
  {Mamon}}{1982}]{mamon}
{Binney} J.,  {Mamon} G.~A.,  1982, \mn@doi [\mnras] {10.1093/mnras/200.2.361},
  \href {http://adsabs.harvard.edu/abs/1982MNRAS.200..361B} {200, 361}

\bibitem[\protect\citeauthoryear{{Binney} \& {Tremaine}}{{Binney} \&
  {Tremaine}}{2008}]{galactic}
{Binney} J.,  {Tremaine} S.,  2008, {Galactic Dynamics: Second Edition}.
Princeton Univ. Press, Princeton, NJ

\bibitem[\protect\citeauthoryear{{Breddels}, {Helmi}, {van den Bosch}, {van de
  Ven}  \& {Battaglia}}{{Breddels} et~al.}{2013}]{breddels}
{Breddels} M.~A.,  {Helmi} A.,  {van den Bosch} R.~C.~E.,  {van de Ven} G.,
  {Battaglia} G.,  2013, \mn@doi [\mnras] {10.1093/mnras/stt956}, \href
  {http://adsabs.harvard.edu/abs/2013MNRAS.433.3173B} {433, 3173}

\bibitem[\protect\citeauthoryear{{Campbell} et~al.,}{{Campbell}
  et~al.}{2016}]{campbell}
{Campbell} D.~J.~R.,  et~al., 2016, preprint, \href
  {http://adsabs.harvard.edu/abs/2016arXiv160304443C} {} (\mn@eprint {arXiv}
  {1603.04443})

\bibitem[\protect\citeauthoryear{{Cole}, {Dehnen}  \& {Wilkinson}}{{Cole}
  et~al.}{2011}]{colecore}
{Cole} D.~R.,  {Dehnen} W.,   {Wilkinson} M.~I.,  2011, \mn@doi [\mnras]
  {10.1111/j.1365-2966.2011.19110.x}, \href
  {http://adsabs.harvard.edu/abs/2011MNRAS.416.1118C} {416, 1118}

\bibitem[\protect\citeauthoryear{{Col{\'{\i}}n}, {Klypin}, {Valenzuela}  \&
  {Gottl{\"o}ber}}{{Col{\'{\i}}n} et~al.}{2004}]{colin}
{Col{\'{\i}}n} P.,  {Klypin} A.,  {Valenzuela} O.,   {Gottl{\"o}ber} S.,  2004,
  \mn@doi [\apj] {10.1086/422463}, \href
  {http://cdsads.u-strasbg.fr/abs/2004ApJ...612...50C} {612, 50}

\bibitem[\protect\citeauthoryear{{Crain} et~al.,}{{Crain}
  et~al.}{2015}]{eagle1}
{Crain} R.~A.,  et~al., 2015, \mn@doi [\mnras] {10.1093/mnras/stv725}, \href
  {http://adsabs.harvard.edu/abs/2015MNRAS.450.1937C} {450, 1937}

\bibitem[\protect\citeauthoryear{{Dalla Vecchia} \& {Schaye}}{{Dalla Vecchia}
  \& {Schaye}}{2012}]{anarchy}
{Dalla Vecchia} C.,  {Schaye} J.,  2012, \mn@doi [\mnras]
  {10.1111/j.1365-2966.2012.21704.x}, \href
  {http://adsabs.harvard.edu/abs/2012MNRAS.426..140D} {426, 140}

\bibitem[\protect\citeauthoryear{{Davis}, {Efstathiou}, {Frenk}  \&
  {White}}{{Davis} et~al.}{1985}]{fof}
{Davis} M.,  {Efstathiou} G.,  {Frenk} C.~S.,   {White} S.~D.~M.,  1985,
  \mn@doi [\apj] {10.1086/163168}, \href
  {http://adsabs.harvard.edu/abs/1985ApJ...292..371D} {292, 371}

\bibitem[\protect\citeauthoryear{{Del Popolo} \& {Kroupa}}{{Del Popolo} \&
  {Kroupa}}{2009}]{delpopolo}
{Del Popolo} A.,  {Kroupa} P.,  2009, \mn@doi [\aap]
  {10.1051/0004-6361/200811404}, \href
  {http://adsabs.harvard.edu/abs/2009A%26A...502..733D} {502, 733}

\bibitem[\protect\citeauthoryear{{Diemand}, {Zemp}, {Moore}, {Stadel}  \&
  {Carollo}}{{Diemand} et~al.}{2005}]{diemand}
{Diemand} J.,  {Zemp} M.,  {Moore} B.,  {Stadel} J.,   {Carollo} C.~M.,  2005,
  \mn@doi [\mnras] {10.1111/j.1365-2966.2005.09601.x}, \href
  {http://adsabs.harvard.edu/abs/2005MNRAS.364..665D} {364, 665}

\bibitem[\protect\citeauthoryear{{Dubinski} \& {Carlberg}}{{Dubinski} \&
  {Carlberg}}{1991}]{dubinski}
{Dubinski} J.,  {Carlberg} R.~G.,  1991, \mn@doi [\apj] {10.1086/170451}, \href
  {http://adsabs.harvard.edu/abs/1991ApJ...378..496D} {378, 496}

\bibitem[\protect\citeauthoryear{{El-Zant}, {Shlosman}  \& {Hoffman}}{{El-Zant}
  et~al.}{2001}]{elzant}
{El-Zant} A.,  {Shlosman} I.,   {Hoffman} Y.,  2001, \mn@doi [\apj]
  {10.1086/322516}, \href {http://adsabs.harvard.edu/abs/2001ApJ...560..636E}
  {560, 636}

\bibitem[\protect\citeauthoryear{{Fattahi} et~al.,}{{Fattahi}
  et~al.}{2016}]{fattahi}
{Fattahi} A.,  et~al., 2016, \mn@doi [\mnras] {10.1093/mnras/stv2970}, \href
  {http://adsabs.harvard.edu/abs/2016MNRAS.457..844F} {457, 844}

\bibitem[\protect\citeauthoryear{{Flores} \& {Primack}}{{Flores} \&
  {Primack}}{1994}]{flores}
{Flores} R.~A.,  {Primack} J.~R.,  1994, \mn@doi [\apjl] {10.1086/187350},
  \href {http://adsabs.harvard.edu/abs/1994ApJ...427L...1F} {427, L1}

\bibitem[\protect\citeauthoryear{Gonz{\'a}lez}{Gonz{\'a}lez}{2009}]{fibonacci}
Gonz{\'a}lez {\'A}.,  2009, \mn@doi [Mathematical Geosciences]
  {10.1007/s11004-009-9257-x}, 42, 49

\bibitem[\protect\citeauthoryear{{Gonzalez-Samaniego}, {Bullock},
  {Boylan-Kolchin}, {Fitts}, {Elbert}, {Hopkins}, {Kere{\v s}}  \&
  {Faucher-Gigu{\`e}re}}{{Gonzalez-Samaniego} et~al.}{2017}]{fireestimator}
{Gonzalez-Samaniego} A.,  {Bullock} J.~S.,  {Boylan-Kolchin} M.,  {Fitts} A.,
  {Elbert} O.~D.,  {Hopkins} P.~F.,  {Kere{\v s}} D.,   {Faucher-Gigu{\`e}re}
  C.-A.,  2017, preprint, \href
  {http://adsabs.harvard.edu/abs/2017arXiv170605383G} {} (\mn@eprint {arXiv}
  {1706.05383})

\bibitem[\protect\citeauthoryear{Hastie, Tibshirani  \& Friedman}{Hastie
  et~al.}{2001}]{gmm}
Hastie T.,  Tibshirani R.,   Friedman J.,  2001, The Elements of Statistical
  Learning.
Springer Series in Statistics, Springer New York Inc., New York, NY, USA

\bibitem[\protect\citeauthoryear{{Hopkins}}{{Hopkins}}{2013}]{hopkins}
{Hopkins} P.~F.,  2013, \mn@doi [\mnras] {10.1093/mnras/sts210}, \href
  {http://adsabs.harvard.edu/abs/2013MNRAS.428.2840H} {428, 2840}

\bibitem[\protect\citeauthoryear{{Jenkins}}{{Jenkins}}{2013}]{dove}
{Jenkins} A.,  2013, \mn@doi [\mnras] {10.1093/mnras/stt1154}, \href
  {http://adsabs.harvard.edu/abs/2013MNRAS.434.2094J} {434, 2094}

\bibitem[\protect\citeauthoryear{{Kawata}, {Arimoto}, {Cen}  \&
  {Gibson}}{{Kawata} et~al.}{2006}]{photoion}
{Kawata} D.,  {Arimoto} N.,  {Cen} R.,   {Gibson} B.~K.,  2006, \mn@doi [\apj]
  {10.1086/500633}, \href {http://adsabs.harvard.edu/abs/2006ApJ...641..785K}
  {641, 785}

\bibitem[\protect\citeauthoryear{{Klypin}, {Kravtsov}, {Bullock}  \&
  {Primack}}{{Klypin} et~al.}{2001}]{klypinslope}
{Klypin} A.,  {Kravtsov} A.~V.,  {Bullock} J.~S.,   {Primack} J.~R.,  2001,
  \mn@doi [\apj] {10.1086/321400}, \href
  {http://adsabs.harvard.edu/abs/2001ApJ...554..903K} {554, 903}

\bibitem[\protect\citeauthoryear{{Komatsu} et~al.,}{{Komatsu}
  et~al.}{2011}]{wmap7}
{Komatsu} E.,  et~al., 2011, \mn@doi [\apjs] {10.1088/0067-0049/192/2/18},
  \href {http://adsabs.harvard.edu/abs/2011ApJS..192...18K} {192, 18}

\bibitem[\protect\citeauthoryear{{Kowalczyk}, {{\L}okas}, {Kazantzidis}  \&
  {Mayer}}{{Kowalczyk} et~al.}{2013}]{orientation}
{Kowalczyk} K.,  {{\L}okas} E.~L.,  {Kazantzidis} S.,   {Mayer} L.,  2013,
  \mn@doi [\mnras] {10.1093/mnras/stt376}, \href
  {http://adsabs.harvard.edu/abs/2013MNRAS.431.2796K} {431, 2796}

\bibitem[\protect\citeauthoryear{{Laporte}, {Walker}  \&
  {Pe{\~n}arrubia}}{{Laporte} et~al.}{2013a}]{triaxality}
{Laporte} C.~F.~P.,  {Walker} M.~G.,   {Pe{\~n}arrubia} J.,  2013a, \mn@doi
  [\mnras] {10.1093/mnrasl/slt057}, \href
  {http://adsabs.harvard.edu/abs/2013MNRAS.433L..54L} {433, L54}

\bibitem[\protect\citeauthoryear{{Laporte}, {White}, {Naab}  \&
  {Gao}}{{Laporte} et~al.}{2013b}]{laporteconstruction}
{Laporte} C.~F.~P.,  {White} S.~D.~M.,  {Naab} T.,   {Gao} L.,  2013b, \mn@doi
  [\mnras] {10.1093/mnras/stt912}, \href
  {http://adsabs.harvard.edu/abs/2013MNRAS.435..901L} {435, 901}

\bibitem[\protect\citeauthoryear{{Mashchenko}, {Wadsley}  \&
  {Couchman}}{{Mashchenko} et~al.}{2008}]{maschenko}
{Mashchenko} S.,  {Wadsley} J.,   {Couchman} H.~M.~P.,  2008, \mn@doi [Science]
  {10.1126/science.1148666}, \href
  {http://adsabs.harvard.edu/abs/2008Sci...319..174M} {319, 174}

\bibitem[\protect\citeauthoryear{{Merritt}}{{Merritt}}{1985}]{merritt}
{Merritt} D.,  1985, \mn@doi [\aj] {10.1086/113810}, \href
  {http://cdsads.u-strasbg.fr/abs/1985AJ.....90.1027M} {90, 1027}

\bibitem[\protect\citeauthoryear{Moore}{Moore}{1994}]{Moorecore}
Moore B.,  1994, \mn@doi [Nature] {10.1038/370629a0}, 370, 629

\bibitem[\protect\citeauthoryear{{Moore}, {Quinn}, {Governato}, {Stadel}  \&
  {Lake}}{{Moore} et~al.}{1999}]{mooreslope}
{Moore} B.,  {Quinn} T.,  {Governato} F.,  {Stadel} J.,   {Lake} G.,  1999,
  \mn@doi [\mnras] {10.1046/j.1365-8711.1999.03039.x}, \href
  {http://adsabs.harvard.edu/abs/1999MNRAS.310.1147M} {310, 1147}

\bibitem[\protect\citeauthoryear{{Navarro}, {Eke}  \& {Frenk}}{{Navarro}
  et~al.}{1996a}]{navarro1996a}
{Navarro} J.~F.,  {Eke} V.~R.,   {Frenk} C.~S.,  1996a, \mn@doi [\mnras]
  {10.1093/mnras/283.3.L72}, \href
  {http://adsabs.harvard.edu/abs/1996MNRAS.283L..72N} {283, L72}

\bibitem[\protect\citeauthoryear{{Navarro}, {Frenk}  \& {White}}{{Navarro}
  et~al.}{1996b}]{navarro1996b}
{Navarro} J.~F.,  {Frenk} C.~S.,   {White} S.~D.~M.,  1996b, \mn@doi [\apj]
  {10.1086/177173}, \href {http://adsabs.harvard.edu/abs/1996ApJ...462..563N}
  {462, 563}

\bibitem[\protect\citeauthoryear{{Navarro}, {Frenk}  \& {White}}{{Navarro}
  et~al.}{1997}]{nfw}
{Navarro} J.~F.,  {Frenk} C.~S.,   {White} S.~D.~M.,  1997, \mn@doi [\apj]
  {10.1086/304888}, \href {http://adsabs.harvard.edu/abs/1997ApJ...490..493N}
  {490, 493}

\bibitem[\protect\citeauthoryear{{Oh} et~al.,}{{Oh}
  et~al.}{2015}]{littlethings}
{Oh} S.-H.,  et~al., 2015, \mn@doi [\aj] {10.1088/0004-6256/149/6/180}, \href
  {http://adsabs.harvard.edu/abs/2015AJ....149..180O} {149, 180}

\bibitem[\protect\citeauthoryear{{Okamoto}, {Eke}, {Frenk}  \&
  {Jenkins}}{{Okamoto} et~al.}{2005}]{metal2}
{Okamoto} T.,  {Eke} V.~R.,  {Frenk} C.~S.,   {Jenkins} A.,  2005, \mn@doi
  [\mnras] {10.1111/j.1365-2966.2005.09525.x}, \href
  {http://adsabs.harvard.edu/abs/2005MNRAS.363.1299O} {363, 1299}

\bibitem[\protect\citeauthoryear{{Okamoto}, {Shimizu}  \& {Yoshida}}{{Okamoto}
  et~al.}{2014}]{metal1}
{Okamoto} T.,  {Shimizu} I.,   {Yoshida} N.,  2014, \mn@doi [\pasj]
  {10.1093/pasj/psu046}, \href
  {http://adsabs.harvard.edu/abs/2014PASJ...66...70O} {66, 70}

\bibitem[\protect\citeauthoryear{{Osipkov}}{{Osipkov}}{1979}]{osipkov}
{Osipkov} L.~P.,  1979, Pisma v Astronomicheskii Zhurnal, \href
  {http://cdsads.u-strasbg.fr/abs/1979PAZh....5...77O} {5, 77}

\bibitem[\protect\citeauthoryear{{Pontzen} \& {Governato}}{{Pontzen} \&
  {Governato}}{2012}]{pontzen}
{Pontzen} A.,  {Governato} F.,  2012, \mn@doi [\mnras]
  {10.1111/j.1365-2966.2012.20571.x}, \href
  {http://adsabs.harvard.edu/abs/2012MNRAS.421.3464P} {421, 3464}

\bibitem[\protect\citeauthoryear{{Power}, {Navarro}, {Jenkins}, {Frenk},
  {White}, {Springel}, {Stadel}  \& {Quinn}}{{Power} et~al.}{2003}]{power}
{Power} C.,  {Navarro} J.~F.,  {Jenkins} A.,  {Frenk} C.~S.,  {White} S.~D.~M.,
   {Springel} V.,  {Stadel} J.,   {Quinn} T.,  2003, \mn@doi [\mnras]
  {10.1046/j.1365-8711.2003.05925.x}, \href
  {http://adsabs.harvard.edu/abs/2003MNRAS.338...14P} {338, 14}

\bibitem[\protect\citeauthoryear{{Pryor} \& {Kormendy}}{{Pryor} \&
  {Kormendy}}{1990}]{pryor}
{Pryor} C.,  {Kormendy} J.,  1990, \mn@doi [\aj] {10.1086/115496}, \href
  {http://adsabs.harvard.edu/abs/1990AJ....100..127P} {100, 127}

\bibitem[\protect\citeauthoryear{{Read} \& {Gilmore}}{{Read} \&
  {Gilmore}}{2005}]{readgilmore}
{Read} J.~I.,  {Gilmore} G.,  2005, \mn@doi [\mnras]
  {10.1111/j.1365-2966.2004.08424.x}, \href
  {http://adsabs.harvard.edu/abs/2005MNRAS.356..107R} {356, 107}

\bibitem[\protect\citeauthoryear{{Richardson} \& {Fairbairn}}{{Richardson} \&
  {Fairbairn}}{2014}]{richardson}
{Richardson} T.,  {Fairbairn} M.,  2014, \mn@doi [\mnras]
  {10.1093/mnras/stu691}, \href
  {http://adsabs.harvard.edu/abs/2014MNRAS.441.1584R} {441, 1584}

\bibitem[\protect\citeauthoryear{{Sales}, {Navarro}, {Theuns}, {Schaye},
  {White}, {Frenk}, {Crain}  \& {Dalla Vecchia}}{{Sales} et~al.}{2012}]{kappa}
{Sales} L.~V.,  {Navarro} J.~F.,  {Theuns} T.,  {Schaye} J.,  {White} S.~D.~M.,
   {Frenk} C.~S.,  {Crain} R.~A.,   {Dalla Vecchia} C.,  2012, \mn@doi [\mnras]
  {10.1111/j.1365-2966.2012.20975.x}, \href
  {http://adsabs.harvard.edu/abs/2012MNRAS.423.1544S} {423, 1544}

\bibitem[\protect\citeauthoryear{{S{\'a}nchez-Salcedo}, {Reyes-Iturbide}  \&
  {Hernandez}}{{S{\'a}nchez-Salcedo} et~al.}{2006}]{sanchez}
{S{\'a}nchez-Salcedo} F.~J.,  {Reyes-Iturbide} J.,   {Hernandez} X.,  2006,
  \mn@doi [\mnras] {10.1111/j.1365-2966.2006.10602.x}, \href
  {http://adsabs.harvard.edu/abs/2006MNRAS.370.1829S} {370, 1829}

\bibitem[\protect\citeauthoryear{{Sawala} et~al.,}{{Sawala}
  et~al.}{2016}]{sawalapuzzles}
{Sawala} T.,  et~al., 2016, \mn@doi [\mnras] {10.1093/mnras/stw145}, \href
  {http://adsabs.harvard.edu/abs/2016MNRAS.457.1931S} {457, 1931}

\bibitem[\protect\citeauthoryear{{Schaller}, {Dalla Vecchia}, {Schaye},
  {Bower}, {Theuns}, {Crain}, {Furlong}  \& {McCarthy}}{{Schaller}
  et~al.}{2015}]{anarchy2}
{Schaller} M.,  {Dalla Vecchia} C.,  {Schaye} J.,  {Bower} R.~G.,  {Theuns} T.,
   {Crain} R.~A.,  {Furlong} M.,   {McCarthy} I.~G.,  2015, \mn@doi [\mnras]
  {10.1093/mnras/stv2169}, \href
  {http://adsabs.harvard.edu/abs/2015MNRAS.454.2277S} {454, 2277}

\bibitem[\protect\citeauthoryear{{Schaye} et~al.,}{{Schaye}
  et~al.}{2015}]{eagle2}
{Schaye} J.,  et~al., 2015, \mn@doi [\mnras] {10.1093/mnras/stu2058}, \href
  {http://adsabs.harvard.edu/abs/2015MNRAS.446..521S} {446, 521}

\bibitem[\protect\citeauthoryear{{Springel}}{{Springel}}{2005}]{gadget}
{Springel} V.,  2005, \mn@doi [\mnras] {10.1111/j.1365-2966.2005.09655.x},
  \href {http://adsabs.harvard.edu/abs/2005MNRAS.364.1105S} {364, 1105}

\bibitem[\protect\citeauthoryear{{Springel}, {White}, {Tormen}  \&
  {Kauffmann}}{{Springel} et~al.}{2001}]{subfind}
{Springel} V.,  {White} S.~D.~M.,  {Tormen} G.,   {Kauffmann} G.,  2001,
  \mn@doi [\mnras] {10.1046/j.1365-8711.2001.04912.x}, \href
  {http://adsabs.harvard.edu/abs/2001MNRAS.328..726S} {328, 726}

\bibitem[\protect\citeauthoryear{{Springel} et~al.,}{{Springel}
  et~al.}{2008}]{aquarius}
{Springel} V.,  et~al., 2008, \mn@doi [\mnras]
  {10.1111/j.1365-2966.2008.14066.x}, \href
  {http://adsabs.harvard.edu/abs/2008MNRAS.391.1685S} {391, 1685}

\bibitem[\protect\citeauthoryear{{Strigari}, {Frenk}  \& {White}}{{Strigari}
  et~al.}{2014}]{strigari}
{Strigari} L.~E.,  {Frenk} C.~S.,   {White} S.~D.~M.,  2014, preprint, \href
  {http://adsabs.harvard.edu/abs/2014arXiv1406.6079S} {} (\mn@eprint {arXiv}
  {1406.6079})

\bibitem[\protect\citeauthoryear{{Tolstoy} et~al.,}{{Tolstoy}
  et~al.}{2004}]{tolstoy}
{Tolstoy} E.,  et~al., 2004, \mn@doi [\apjl] {10.1086/427388}, \href
  {http://adsabs.harvard.edu/abs/2004ApJ...617L.119T} {617, L119}

\bibitem[\protect\citeauthoryear{{Walker} \& {Pe{\~n}arrubia}}{{Walker} \&
  {Pe{\~n}arrubia}}{2011}]{walk}
{Walker} M.~G.,  {Pe{\~n}arrubia} J.,  2011, \mn@doi [\apj]
  {10.1088/0004-637X/742/1/20}, \href
  {http://adsabs.harvard.edu/abs/2011ApJ...742...20W} {742, 20}

\bibitem[\protect\citeauthoryear{{Walker}, {Mateo}, {Olszewski},
  {Pe{\~n}arrubia}, {Wyn Evans}  \& {Gilmore}}{{Walker}
  et~al.}{2009}]{walkerest}
{Walker} M.~G.,  {Mateo} M.,  {Olszewski} E.~W.,  {Pe{\~n}arrubia} J.,  {Wyn
  Evans} N.,   {Gilmore} G.,  2009, \mn@doi [\apj]
  {10.1088/0004-637X/704/2/1274}, \href
  {http://adsabs.harvard.edu/abs/2009ApJ...704.1274W} {704, 1274}

\bibitem[\protect\citeauthoryear{{Wolf}, {Martinez}, {Bullock}, {Kaplinghat},
  {Geha}, {Mu{\~n}oz}, {Simon}  \& {Avedo}}{{Wolf} et~al.}{2010}]{wolf}
{Wolf} J.,  {Martinez} G.~D.,  {Bullock} J.~S.,  {Kaplinghat} M.,  {Geha} M.,
  {Mu{\~n}oz} R.~R.,  {Simon} J.~D.,   {Avedo} F.~F.,  2010, \mn@doi [\mnras]
  {10.1111/j.1365-2966.2010.16753.x}, \href
  {http://adsabs.harvard.edu/abs/2010MNRAS.406.1220W} {406, 1220}

\makeatother
\end{thebibliography}


\begin{thebibliography}{}
\makeatletter
\relax
\def\mn@urlcharsother{\let\do\@makeother \do\$\do\&\do\#\do\^\do\_\do\%\do\~}
\def\mn@doi{\begingroup\mn@urlcharsother \@ifnextchar [ {\mn@doi@}
  {\mn@doi@[]}}
\def\mn@doi@[#1]#2{\def\@tempa{#1}\ifx\@tempa\@empty \href
  {http://dx.doi.org/#2} {doi:#2}\else \href {http://dx.doi.org/#2} {#1}\fi
  \endgroup}
\def\mn@eprint#1#2{\mn@eprint@#1:#2::\@nil}
\def\mn@eprint@arXiv#1{\href {http://arxiv.org/abs/#1} {{\tt arXiv:#1}}}
\def\mn@eprint@dblp#1{\href {http://dblp.uni-trier.de/rec/bibtex/#1.xml}
  {dblp:#1}}
\def\mn@eprint@#1:#2:#3:#4\@nil{\def\@tempa {#1}\def\@tempb {#2}\def\@tempc
  {#3}\ifx \@tempc \@empty \let \@tempc \@tempb \let \@tempb \@tempa \fi \ifx
  \@tempb \@empty \def\@tempb {arXiv}\fi \@ifundefined
  {mn@eprint@\@tempb}{\@tempb:\@tempc}{\expandafter \expandafter \csname
  mn@eprint@\@tempb\endcsname \expandafter{\@tempc}}}

\bibitem[\protect\citeauthoryear{{Adams} et~al.,}{{Adams}
  et~al.}{2014}]{adamsgas}
{Adams} J.~J.,  et~al., 2014, \mn@doi [\apj] {10.1088/0004-637X/789/1/63},
  \href {http://adsabs.harvard.edu/abs/2014ApJ...789...63A} {789, 63}

\bibitem[\protect\citeauthoryear{{Agnello} \& {Evans}}{{Agnello} \&
  {Evans}}{2012}]{agnello}
{Agnello} A.,  {Evans} N.~W.,  2012, \mn@doi [\apjl]
  {10.1088/2041-8205/754/2/L39}, \href
  {http://adsabs.harvard.edu/abs/2012ApJ...754L..39A} {754, L39}

\bibitem[\protect\citeauthoryear{Akaike}{Akaike}{1998}]{Akaike}
Akaike H.,  1998, Information Theory and an Extension of the Maximum Likelihood
  Principle.
Springer New York, New York, NY, pp 199--213,
  \mn@doi{10.1007/978-1-4612-1694-0_15}

\bibitem[\protect\citeauthoryear{{Amorisco} \& {Evans}}{{Amorisco} \&
  {Evans}}{2012}]{amorisco}
{Amorisco} N.~C.,  {Evans} N.~W.,  2012, \mn@doi [\mnras]
  {10.1111/j.1365-2966.2011.19684.x}, \href
  {http://adsabs.harvard.edu/abs/2012MNRAS.419..184A} {419, 184}

\bibitem[\protect\citeauthoryear{{Barber}, {Starkenburg}, {Navarro}  \&
  {McConnachie}}{{Barber} et~al.}{2015}]{barber}
{Barber} C.,  {Starkenburg} E.,  {Navarro} J.~F.,   {McConnachie} A.~W.,  2015,
  \mn@doi [\mnras] {10.1093/mnras/stu2494}, \href
  {http://adsabs.harvard.edu/abs/2015MNRAS.447.1112B} {447, 1112}

\bibitem[\protect\citeauthoryear{{Battaglia} et~al.,}{{Battaglia}
  et~al.}{2006}]{battaglia06}
{Battaglia} G.,  et~al., 2006, \mn@doi [\aap] {10.1051/0004-6361:20065720},
  \href {http://adsabs.harvard.edu/abs/2006A%26A...459..423B} {459, 423}

\bibitem[\protect\citeauthoryear{{Battaglia}, {Helmi}, {Tolstoy}, {Irwin},
  {Hill}  \& {Jablonka}}{{Battaglia} et~al.}{2008}]{battaglia}
{Battaglia} G.,  {Helmi} A.,  {Tolstoy} E.,  {Irwin} M.,  {Hill} V.,
  {Jablonka} P.,  2008, \mn@doi [\apjl] {10.1086/590179}, \href
  {http://adsabs.harvard.edu/abs/2008ApJ...681L..13B} {681, L13}

\bibitem[\protect\citeauthoryear{{Battaglia}, {Tolstoy}, {Helmi}, {Irwin},
  {Parisi}, {Hill}  \& {Jablonka}}{{Battaglia} et~al.}{2011}]{battagliasextans}
{Battaglia} G.,  {Tolstoy} E.,  {Helmi} A.,  {Irwin} M.,  {Parisi} P.,  {Hill}
  V.,   {Jablonka} P.,  2011, \mn@doi [\mnras]
  {10.1111/j.1365-2966.2010.17745.x}, \href
  {http://adsabs.harvard.edu/abs/2011MNRAS.411.1013B} {411, 1013}

\bibitem[\protect\citeauthoryear{{Ben{\'{\i}}tez-Llambay}, {Navarro}, {Abadi},
  {Gottl{\"o}ber}, {Yepes}, {Hoffman}  \& {Steinmetz}}{{Ben{\'{\i}}tez-Llambay}
  et~al.}{2016}]{mergers}
{Ben{\'{\i}}tez-Llambay} A.,  {Navarro} J.~F.,  {Abadi} M.~G.,  {Gottl{\"o}ber}
  S.,  {Yepes} G.,  {Hoffman} Y.,   {Steinmetz} M.,  2016, \mn@doi [\mnras]
  {10.1093/mnras/stv2722}, \href
  {http://adsabs.harvard.edu/abs/2016MNRAS.456.1185B} {456, 1185}

\bibitem[\protect\citeauthoryear{{Bett}, {Eke}, {Frenk}, {Jenkins}, {Helly}  \&
  {Navarro}}{{Bett} et~al.}{2007}]{sphericity}
{Bett} P.,  {Eke} V.,  {Frenk} C.~S.,  {Jenkins} A.,  {Helly} J.,   {Navarro}
  J.,  2007, \mn@doi [\mnras] {10.1111/j.1365-2966.2007.11432.x}, \href
  {http://adsabs.harvard.edu/abs/2007MNRAS.376..215B} {376, 215}

\bibitem[\protect\citeauthoryear{{Booth} \& {Schaye}}{{Booth} \&
  {Schaye}}{2009}]{agnbooth}
{Booth} C.~M.,  {Schaye} J.,  2009, \mn@doi [\mnras]
  {10.1111/j.1365-2966.2009.15043.x}, \href
  {http://adsabs.harvard.edu/abs/2009MNRAS.398...53B} {398, 53}

\bibitem[\protect\citeauthoryear{{Breddels}, {Helmi}, {van den Bosch}, {van de
  Ven}  \& {Battaglia}}{{Breddels} et~al.}{2013}]{breddels}
{Breddels} M.~A.,  {Helmi} A.,  {van den Bosch} R.~C.~E.,  {van de Ven} G.,
  {Battaglia} G.,  2013, \mn@doi [\mnras] {10.1093/mnras/stt956}, \href
  {http://adsabs.harvard.edu/abs/2013MNRAS.433.3173B} {433, 3173}

\bibitem[\protect\citeauthoryear{{Brooks} \& {Zolotov}}{{Brooks} \&
  {Zolotov}}{2014}]{brooks2014}
{Brooks} A.~M.,  {Zolotov} A.,  2014, \mn@doi [\apj]
  {10.1088/0004-637X/786/2/87}, \href
  {http://adsabs.harvard.edu/abs/2014ApJ...786...87B} {786, 87}

\bibitem[\protect\citeauthoryear{{Campbell} et~al.,}{{Campbell}
  et~al.}{2017}]{campbell}
{Campbell} D.~J.~R.,  et~al., 2017, \mn@doi [\mnras] {10.1093/mnras/stx975},
  \href {http://adsabs.harvard.edu/abs/2017MNRAS.469.2335C} {469, 2335}

\bibitem[\protect\citeauthoryear{{Cole}, {Dehnen}  \& {Wilkinson}}{{Cole}
  et~al.}{2011}]{colecore}
{Cole} D.~R.,  {Dehnen} W.,   {Wilkinson} M.~I.,  2011, \mn@doi [\mnras]
  {10.1111/j.1365-2966.2011.19110.x}, \href
  {http://adsabs.harvard.edu/abs/2011MNRAS.416.1118C} {416, 1118}

\bibitem[\protect\citeauthoryear{{Crain} et~al.,}{{Crain}
  et~al.}{2015}]{eagle1}
{Crain} R.~A.,  et~al., 2015, \mn@doi [\mnras] {10.1093/mnras/stv725}, \href
  {http://adsabs.harvard.edu/abs/2015MNRAS.450.1937C} {450, 1937}

\bibitem[\protect\citeauthoryear{{Dalla Vecchia} \& {Schaye}}{{Dalla Vecchia}
  \& {Schaye}}{2012}]{anarchy}
{Dalla Vecchia} C.,  {Schaye} J.,  2012, \mn@doi [\mnras]
  {10.1111/j.1365-2966.2012.21704.x}, \href
  {http://adsabs.harvard.edu/abs/2012MNRAS.426..140D} {426, 140}

\bibitem[\protect\citeauthoryear{{Davis}, {Efstathiou}, {Frenk}  \&
  {White}}{{Davis} et~al.}{1985}]{fof}
{Davis} M.,  {Efstathiou} G.,  {Frenk} C.~S.,   {White} S.~D.~M.,  1985,
  \mn@doi [\apj] {10.1086/163168}, \href
  {http://adsabs.harvard.edu/abs/1985ApJ...292..371D} {292, 371}

\bibitem[\protect\citeauthoryear{{Del Popolo} \& {Kroupa}}{{Del Popolo} \&
  {Kroupa}}{2009}]{delpopolo}
{Del Popolo} A.,  {Kroupa} P.,  2009, \mn@doi [\aap]
  {10.1051/0004-6361/200811404}, \href
  {http://adsabs.harvard.edu/abs/2009A%26A...502..733D} {502, 733}

\bibitem[\protect\citeauthoryear{{El-Zant}, {Shlosman}  \& {Hoffman}}{{El-Zant}
  et~al.}{2001}]{elzant}
{El-Zant} A.,  {Shlosman} I.,   {Hoffman} Y.,  2001, \mn@doi [\apj]
  {10.1086/322516}, \href {http://adsabs.harvard.edu/abs/2001ApJ...560..636E}
  {560, 636}

\bibitem[\protect\citeauthoryear{{Fattahi} et~al.,}{{Fattahi}
  et~al.}{2016}]{fattahi}
{Fattahi} A.,  et~al., 2016, \mn@doi [\mnras] {10.1093/mnras/stv2970}, \href
  {http://adsabs.harvard.edu/abs/2016MNRAS.457..844F} {457, 844}

\bibitem[\protect\citeauthoryear{{Flores} \& {Primack}}{{Flores} \&
  {Primack}}{1994}]{flores}
{Flores} R.~A.,  {Primack} J.~R.,  1994, \mn@doi [\apjl] {10.1086/187350},
  \href {http://adsabs.harvard.edu/abs/1994ApJ...427L...1F} {427, L1}

\bibitem[\protect\citeauthoryear{{Gaia Collaboration} et~al.,}{{Gaia
  Collaboration} et~al.}{2016}]{gaia}
{Gaia Collaboration} et~al., 2016, \mn@doi [\aap]
  {10.1051/0004-6361/201629272}, \href
  {http://adsabs.harvard.edu/abs/2016A%26A...595A...1G} {595, A1}

\bibitem[\protect\citeauthoryear{{Gilmore}, {Wilkinson}, {Wyse}, {Kleyna},
  {Koch}, {Evans}  \& {Grebel}}{{Gilmore} et~al.}{2007}]{gilmore2007}
{Gilmore} G.,  {Wilkinson} M.~I.,  {Wyse} R.~F.~G.,  {Kleyna} J.~T.,  {Koch}
  A.,  {Evans} N.~W.,   {Grebel} E.~K.,  2007, \mn@doi [\apj] {10.1086/518025},
  \href {http://adsabs.harvard.edu/abs/2007ApJ...663..948G} {663, 948}

\bibitem[\protect\citeauthoryear{Gonz{\'a}lez}{Gonz{\'a}lez}{2009}]{fibonacci}
Gonz{\'a}lez {\'A}.,  2009, \mn@doi [Mathematical Geosciences]
  {10.1007/s11004-009-9257-x}, 42, 49

\bibitem[\protect\citeauthoryear{{Gonz{\'a}lez-Samaniego}, {Bullock},
  {Boylan-Kolchin}, {Fitts}, {Elbert}, {Hopkins}, {Kere{\v s}}  \&
  {Faucher-Gigu{\`e}re}}{{Gonz{\'a}lez-Samaniego} et~al.}{2017}]{fireestimator}
{Gonz{\'a}lez-Samaniego} A.,  {Bullock} J.~S.,  {Boylan-Kolchin} M.,  {Fitts}
  A.,  {Elbert} O.~D.,  {Hopkins} P.~F.,  {Kere{\v s}} D.,
  {Faucher-Gigu{\`e}re} C.-A.,  2017, \mn@doi [\mnras] {10.1093/mnras/stx2322},
  \href {http://adsabs.harvard.edu/abs/2017MNRAS.472.4786G} {472, 4786}

\bibitem[\protect\citeauthoryear{Hastie, Tibshirani  \& Friedman}{Hastie
  et~al.}{2001}]{gmm}
Hastie T.,  Tibshirani R.,   Friedman J.,  2001, The Elements of Statistical
  Learning.
Springer Series in Statistics, Springer New York Inc., New York, NY, USA

\bibitem[\protect\citeauthoryear{{Hopkins}}{{Hopkins}}{2013}]{hopkins}
{Hopkins} P.~F.,  2013, \mn@doi [\mnras] {10.1093/mnras/sts210}, \href
  {http://adsabs.harvard.edu/abs/2013MNRAS.428.2840H} {428, 2840}

\bibitem[\protect\citeauthoryear{{Jenkins}}{{Jenkins}}{2013}]{dove}
{Jenkins} A.,  2013, \mn@doi [\mnras] {10.1093/mnras/stt1154}, \href
  {http://adsabs.harvard.edu/abs/2013MNRAS.434.2094J} {434, 2094}

\bibitem[\protect\citeauthoryear{{Jin}, {Helmi}  \& {Breddels}}{{Jin}
  et~al.}{2015}]{sculptorproper}
{Jin} S.,  {Helmi} A.,   {Breddels} M.,  2015, preprint, \href
  {http://adsabs.harvard.edu/abs/2015arXiv150201215J} {} (\mn@eprint {arXiv}
  {1502.01215})

\bibitem[\protect\citeauthoryear{{Kawata}, {Arimoto}, {Cen}  \&
  {Gibson}}{{Kawata} et~al.}{2006}]{photoion}
{Kawata} D.,  {Arimoto} N.,  {Cen} R.,   {Gibson} B.~K.,  2006, \mn@doi [\apj]
  {10.1086/500633}, \href {http://adsabs.harvard.edu/abs/2006ApJ...641..785K}
  {641, 785}

\bibitem[\protect\citeauthoryear{{Komatsu} et~al.,}{{Komatsu}
  et~al.}{2011}]{wmap7}
{Komatsu} E.,  et~al., 2011, \mn@doi [\apjs] {10.1088/0067-0049/192/2/18},
  \href {http://adsabs.harvard.edu/abs/2011ApJS..192...18K} {192, 18}

\bibitem[\protect\citeauthoryear{{Kowalczyk}, {{\L}okas}, {Kazantzidis}  \&
  {Mayer}}{{Kowalczyk} et~al.}{2013}]{orientation}
{Kowalczyk} K.,  {{\L}okas} E.~L.,  {Kazantzidis} S.,   {Mayer} L.,  2013,
  \mn@doi [\mnras] {10.1093/mnras/stt376}, \href
  {http://adsabs.harvard.edu/abs/2013MNRAS.431.2796K} {431, 2796}

\bibitem[\protect\citeauthoryear{{Kuzio de Naray}, {McGaugh}, {de Blok}  \&
  {Bosma}}{{Kuzio de Naray} et~al.}{2006}]{kuzio}
{Kuzio de Naray} R.,  {McGaugh} S.~S.,  {de Blok} W.~J.~G.,   {Bosma} A.,
  2006, \mn@doi [\apjs] {10.1086/505345}, \href
  {http://adsabs.harvard.edu/abs/2006ApJS..165..461K} {165, 461}

\bibitem[\protect\citeauthoryear{{Laporte}, {Walker}  \&
  {Pe{\~n}arrubia}}{{Laporte} et~al.}{2013a}]{triaxality}
{Laporte} C.~F.~P.,  {Walker} M.~G.,   {Pe{\~n}arrubia} J.,  2013a, \mn@doi
  [\mnras] {10.1093/mnrasl/slt057}, \href
  {http://adsabs.harvard.edu/abs/2013MNRAS.433L..54L} {433, L54}

\bibitem[\protect\citeauthoryear{{Laporte}, {White}, {Naab}  \&
  {Gao}}{{Laporte} et~al.}{2013b}]{laporteconstruction}
{Laporte} C.~F.~P.,  {White} S.~D.~M.,  {Naab} T.,   {Gao} L.,  2013b, \mn@doi
  [\mnras] {10.1093/mnras/stt912}, \href
  {http://adsabs.harvard.edu/abs/2013MNRAS.435..901L} {435, 901}

\bibitem[\protect\citeauthoryear{{Mashchenko}, {Wadsley}  \&
  {Couchman}}{{Mashchenko} et~al.}{2008}]{maschenko}
{Mashchenko} S.,  {Wadsley} J.,   {Couchman} H.~M.~P.,  2008, \mn@doi [Science]
  {10.1126/science.1148666}, \href
  {http://adsabs.harvard.edu/abs/2008Sci...319..174M} {319, 174}

\bibitem[\protect\citeauthoryear{Moore}{Moore}{1994}]{Moorecore}
Moore B.,  1994, \mn@doi [Nature] {10.1038/370629a0}, 370, 629

\bibitem[\protect\citeauthoryear{{Navarro}, {Eke}  \& {Frenk}}{{Navarro}
  et~al.}{1996}]{navarro1996b}
{Navarro} J.~F.,  {Eke} V.~R.,   {Frenk} C.~S.,  1996, \mn@doi [\mnras]
  {10.1093/mnras/283.3.L72}, \href
  {http://adsabs.harvard.edu/abs/1996MNRAS.283L..72N} {283, L72}

\bibitem[\protect\citeauthoryear{{Navarro}, {Frenk}  \& {White}}{{Navarro}
  et~al.}{1997}]{nfw}
{Navarro} J.~F.,  {Frenk} C.~S.,   {White} S.~D.~M.,  1997, \mn@doi [\apj]
  {10.1086/304888}, \href {http://adsabs.harvard.edu/abs/1997ApJ...490..493N}
  {490, 493}

\bibitem[\protect\citeauthoryear{{Oh}, {Brook}, {Governato}, {Brinks}, {Mayer},
  {de Blok}, {Brooks}  \& {Walter}}{{Oh} et~al.}{2011}]{ohthings}
{Oh} S.~H.,  {Brook} C.,  {Governato} F.,  {Brinks} E.,  {Mayer} L.,  {de Blok}
  W.~J.~G.,  {Brooks} A.,   {Walter} F.,  2011, \mn@doi [\aj]
  {10.1088/0004-6256/142/1/24}, \href
  {http://adsabs.harvard.edu/abs/2011AJ....142...24O} {142, 24}

\bibitem[\protect\citeauthoryear{{Oh} et~al.,}{{Oh}
  et~al.}{2015}]{littlethings}
{Oh} S.~H.,  et~al., 2015, \mn@doi [\aj] {10.1088/0004-6256/149/6/180}, \href
  {http://adsabs.harvard.edu/abs/2015AJ....149..180O} {149, 180}

\bibitem[\protect\citeauthoryear{{Okamoto}, {Eke}, {Frenk}  \&
  {Jenkins}}{{Okamoto} et~al.}{2005}]{metal2}
{Okamoto} T.,  {Eke} V.~R.,  {Frenk} C.~S.,   {Jenkins} A.,  2005, \mn@doi
  [\mnras] {10.1111/j.1365-2966.2005.09525.x}, \href
  {http://adsabs.harvard.edu/abs/2005MNRAS.363.1299O} {363, 1299}

\bibitem[\protect\citeauthoryear{{Okamoto}, {Shimizu}  \& {Yoshida}}{{Okamoto}
  et~al.}{2014}]{metal1}
{Okamoto} T.,  {Shimizu} I.,   {Yoshida} N.,  2014, \mn@doi [\pasj]
  {10.1093/pasj/psu046}, \href
  {http://adsabs.harvard.edu/abs/2014PASJ...66...70O} {66, 70}

\bibitem[\protect\citeauthoryear{{Oman}, {Marasco}, {Navarro}, {Frenk},
  {Schaye}  \& {Ben{\'{\i}}tez-Llambay}}{{Oman} et~al.}{2017}]{kylecurves}
{Oman} K.~A.,  {Marasco} A.,  {Navarro} J.~F.,  {Frenk} C.~S.,  {Schaye} J.,
  {Ben{\'{\i}}tez-Llambay} A.,  2017, preprint, \href
  {http://adsabs.harvard.edu/abs/2017arXiv170607478O} {} (\mn@eprint {arXiv}
  {1706.07478})

\bibitem[\protect\citeauthoryear{{Pineda}, {Hayward}, {Springel}  \& {Mendes de
  Oliveira}}{{Pineda} et~al.}{2017}]{fatalattraction}
{Pineda} J.~C.~B.,  {Hayward} C.~C.,  {Springel} V.,   {Mendes de Oliveira} C.,
   2017, \mn@doi [\mnras] {10.1093/mnras/stw3004}, \href
  {http://adsabs.harvard.edu/abs/2017MNRAS.466...63P} {466, 63}

\bibitem[\protect\citeauthoryear{{Pontzen} \& {Governato}}{{Pontzen} \&
  {Governato}}{2012}]{pontzen}
{Pontzen} A.,  {Governato} F.,  2012, \mn@doi [\mnras]
  {10.1111/j.1365-2966.2012.20571.x}, \href
  {http://adsabs.harvard.edu/abs/2012MNRAS.421.3464P} {421, 3464}

\bibitem[\protect\citeauthoryear{{Power}, {Navarro}, {Jenkins}, {Frenk},
  {White}, {Springel}, {Stadel}  \& {Quinn}}{{Power} et~al.}{2003}]{power}
{Power} C.,  {Navarro} J.~F.,  {Jenkins} A.,  {Frenk} C.~S.,  {White} S.~D.~M.,
   {Springel} V.,  {Stadel} J.,   {Quinn} T.,  2003, \mn@doi [\mnras]
  {10.1046/j.1365-8711.2003.05925.x}, \href
  {http://adsabs.harvard.edu/abs/2003MNRAS.338...14P} {338, 14}

\bibitem[\protect\citeauthoryear{{Pryor} \& {Kormendy}}{{Pryor} \&
  {Kormendy}}{1990}]{pryor}
{Pryor} C.,  {Kormendy} J.,  1990, \mn@doi [\aj] {10.1086/115496}, \href
  {http://adsabs.harvard.edu/abs/1990AJ....100..127P} {100, 127}

\bibitem[\protect\citeauthoryear{{Read} \& {Gilmore}}{{Read} \&
  {Gilmore}}{2005}]{readgilmore}
{Read} J.~I.,  {Gilmore} G.,  2005, \mn@doi [\mnras]
  {10.1111/j.1365-2966.2004.08424.x}, \href
  {http://adsabs.harvard.edu/abs/2005MNRAS.356..107R} {356, 107}

\bibitem[\protect\citeauthoryear{{Richardson} \& {Fairbairn}}{{Richardson} \&
  {Fairbairn}}{2014}]{richardson}
{Richardson} T.,  {Fairbairn} M.,  2014, \mn@doi [\mnras]
  {10.1093/mnras/stu691}, \href
  {http://adsabs.harvard.edu/abs/2014MNRAS.441.1584R} {441, 1584}

\bibitem[\protect\citeauthoryear{{Rosas-Guevara} et~al.,}{{Rosas-Guevara}
  et~al.}{2015}]{accretionmergers}
{Rosas-Guevara} Y.~M.,  et~al., 2015, \mn@doi [\mnras] {10.1093/mnras/stv2056},
  \href {http://adsabs.harvard.edu/abs/2015MNRAS.454.1038R} {454, 1038}

\bibitem[\protect\citeauthoryear{{Sales}, {Navarro}, {Theuns}, {Schaye},
  {White}, {Frenk}, {Crain}  \& {Dalla Vecchia}}{{Sales} et~al.}{2012}]{kappa}
{Sales} L.~V.,  {Navarro} J.~F.,  {Theuns} T.,  {Schaye} J.,  {White} S.~D.~M.,
   {Frenk} C.~S.,  {Crain} R.~A.,   {Dalla Vecchia} C.,  2012, \mn@doi [\mnras]
  {10.1111/j.1365-2966.2012.20975.x}, \href
  {http://adsabs.harvard.edu/abs/2012MNRAS.423.1544S} {423, 1544}

\bibitem[\protect\citeauthoryear{{S{\'a}nchez-Salcedo}, {Reyes-Iturbide}  \&
  {Hernandez}}{{S{\'a}nchez-Salcedo} et~al.}{2006}]{sanchez}
{S{\'a}nchez-Salcedo} F.~J.,  {Reyes-Iturbide} J.,   {Hernandez} X.,  2006,
  \mn@doi [\mnras] {10.1111/j.1365-2966.2006.10602.x}, \href
  {http://adsabs.harvard.edu/abs/2006MNRAS.370.1829S} {370, 1829}

\bibitem[\protect\citeauthoryear{{Sawala} et~al.,}{{Sawala}
  et~al.}{2016}]{sawalapuzzles}
{Sawala} T.,  et~al., 2016, \mn@doi [\mnras] {10.1093/mnras/stw145}, \href
  {http://adsabs.harvard.edu/abs/2016MNRAS.457.1931S} {457, 1931}

\bibitem[\protect\citeauthoryear{{Schaller}, {Dalla Vecchia}, {Schaye},
  {Bower}, {Theuns}, {Crain}, {Furlong}  \& {McCarthy}}{{Schaller}
  et~al.}{2015}]{anarchy2}
{Schaller} M.,  {Dalla Vecchia} C.,  {Schaye} J.,  {Bower} R.~G.,  {Theuns} T.,
   {Crain} R.~A.,  {Furlong} M.,   {McCarthy} I.~G.,  2015, \mn@doi [\mnras]
  {10.1093/mnras/stv2169}, \href
  {http://adsabs.harvard.edu/abs/2015MNRAS.454.2277S} {454, 2277}

\bibitem[\protect\citeauthoryear{{Schaye}}{{Schaye}}{2004}]{starformation}
{Schaye} J.,  2004, \mn@doi [\apj] {10.1086/421232}, \href
  {http://adsabs.harvard.edu/abs/2004ApJ...609..667S} {609, 667}

\bibitem[\protect\citeauthoryear{{Schaye} \& {Dalla Vecchia}}{{Schaye} \&
  {Dalla Vecchia}}{2008}]{starformation2}
{Schaye} J.,  {Dalla Vecchia} C.,  2008, \mn@doi [\mnras]
  {10.1111/j.1365-2966.2007.12639.x}, \href
  {http://adsabs.harvard.edu/abs/2008MNRAS.383.1210S} {383, 1210}

\bibitem[\protect\citeauthoryear{{Schaye} et~al.,}{{Schaye}
  et~al.}{2015}]{eagle2}
{Schaye} J.,  et~al., 2015, \mn@doi [\mnras] {10.1093/mnras/stu2058}, \href
  {http://adsabs.harvard.edu/abs/2015MNRAS.446..521S} {446, 521}

\bibitem[\protect\citeauthoryear{{Spergel} \& {Steinhardt}}{{Spergel} \&
  {Steinhardt}}{2000}]{selfintdm}
{Spergel} D.~N.,  {Steinhardt} P.~J.,  2000, \mn@doi [Physical Review Letters]
  {10.1103/PhysRevLett.84.3760}, \href
  {http://adsabs.harvard.edu/abs/2000PhRvL..84.3760S} {84, 3760}

\bibitem[\protect\citeauthoryear{{Springel}}{{Springel}}{2005}]{gadget}
{Springel} V.,  2005, \mn@doi [\mnras] {10.1111/j.1365-2966.2005.09655.x},
  \href {http://adsabs.harvard.edu/abs/2005MNRAS.364.1105S} {364, 1105}

\bibitem[\protect\citeauthoryear{{Springel}, {White}, {Tormen}  \&
  {Kauffmann}}{{Springel} et~al.}{2001}]{subfind}
{Springel} V.,  {White} S.~D.~M.,  {Tormen} G.,   {Kauffmann} G.,  2001,
  \mn@doi [\mnras] {10.1046/j.1365-8711.2001.04912.x}, \href
  {http://adsabs.harvard.edu/abs/2001MNRAS.328..726S} {328, 726}

\bibitem[\protect\citeauthoryear{{Springel} et~al.,}{{Springel}
  et~al.}{2008}]{aquarius}
{Springel} V.,  et~al., 2008, \mn@doi [\mnras]
  {10.1111/j.1365-2966.2008.14066.x}, \href
  {http://adsabs.harvard.edu/abs/2008MNRAS.391.1685S} {391, 1685}

\bibitem[\protect\citeauthoryear{{Strigari}, {Frenk}  \& {White}}{{Strigari}
  et~al.}{2010}]{strigari2010}
{Strigari} L.~E.,  {Frenk} C.~S.,   {White} S.~D.~M.,  2010, \mn@doi [\mnras]
  {10.1111/j.1365-2966.2010.17287.x}, \href
  {http://adsabs.harvard.edu/abs/2010MNRAS.408.2364S} {408, 2364}

\bibitem[\protect\citeauthoryear{{Strigari}, {Frenk}  \& {White}}{{Strigari}
  et~al.}{2014}]{strigari}
{Strigari} L.~E.,  {Frenk} C.~S.,   {White} S.~D.~M.,  2014, preprint, \href
  {http://adsabs.harvard.edu/abs/2014arXiv1406.6079S} {} (\mn@eprint {arXiv}
  {1406.6079})

\bibitem[\protect\citeauthoryear{{Tolstoy} et~al.,}{{Tolstoy}
  et~al.}{2004}]{tolstoy}
{Tolstoy} E.,  et~al., 2004, \mn@doi [\apjl] {10.1086/427388}, \href
  {http://adsabs.harvard.edu/abs/2004ApJ...617L.119T} {617, L119}

\bibitem[\protect\citeauthoryear{{Walker} \& {Pe{\~n}arrubia}}{{Walker} \&
  {Pe{\~n}arrubia}}{2011}]{walk}
{Walker} M.~G.,  {Pe{\~n}arrubia} J.,  2011, \mn@doi [\apj]
  {10.1088/0004-637X/742/1/20}, \href
  {http://adsabs.harvard.edu/abs/2011ApJ...742...20W} {742, 20}

\bibitem[\protect\citeauthoryear{{Walker}, {Mateo}, {Olszewski},
  {Pe{\~n}arrubia}, {Wyn Evans}  \& {Gilmore}}{{Walker}
  et~al.}{2009}]{walkerest}
{Walker} M.~G.,  {Mateo} M.,  {Olszewski} E.~W.,  {Pe{\~n}arrubia} J.,  {Wyn
  Evans} N.,   {Gilmore} G.,  2009, \mn@doi [\apj]
  {10.1088/0004-637X/704/2/1274}, \href
  {http://adsabs.harvard.edu/abs/2009ApJ...704.1274W} {704, 1274}

\bibitem[\protect\citeauthoryear{{Weinberg} \& {Katz}}{{Weinberg} \&
  {Katz}}{2002}]{weinberg_katz2002}
{Weinberg} M.~D.,  {Katz} N.,  2002, \mn@doi [\apj] {10.1086/343847}, \href
  {http://adsabs.harvard.edu/abs/2002ApJ...580..627W} {580, 627}

\bibitem[\protect\citeauthoryear{{Wiersma}, {Schaye}  \& {Smith}}{{Wiersma}
  et~al.}{2009a}]{gascooling}
{Wiersma} R.~P.~C.,  {Schaye} J.,   {Smith} B.~D.,  2009a, \mn@doi [\mnras]
  {10.1111/j.1365-2966.2008.14191.x}, \href
  {http://adsabs.harvard.edu/abs/2009MNRAS.393...99W} {393, 99}

\bibitem[\protect\citeauthoryear{{Wiersma}, {Schaye}, {Theuns}, {Dalla Vecchia}
   \& {Tornatore}}{{Wiersma} et~al.}{2009b}]{enrichment}
{Wiersma} R.~P.~C.,  {Schaye} J.,  {Theuns} T.,  {Dalla Vecchia} C.,
  {Tornatore} L.,  2009b, \mn@doi [\mnras] {10.1111/j.1365-2966.2009.15331.x},
  \href {http://adsabs.harvard.edu/abs/2009MNRAS.399..574W} {399, 574}

\bibitem[\protect\citeauthoryear{{Wolf}, {Martinez}, {Bullock}, {Kaplinghat},
  {Geha}, {Mu{\~n}oz}, {Simon}  \& {Avedo}}{{Wolf} et~al.}{2010}]{wolf}
{Wolf} J.,  {Martinez} G.~D.,  {Bullock} J.~S.,  {Kaplinghat} M.,  {Geha} M.,
  {Mu{\~n}oz} R.~R.,  {Simon} J.~D.,   {Avedo} F.~F.,  2010, \mn@doi [\mnras]
  {10.1111/j.1365-2966.2010.16753.x}, \href
  {http://adsabs.harvard.edu/abs/2010MNRAS.406.1220W} {406, 1220}

\makeatother
\end{thebibliography}
\bsp	

\appendix
\renewcommand*{\thefigure}{A\arabic{figure}}

\section*{Appendix A}
\label{appendix}

\renewcommand*{\thesubsection}{A\arabic{subsection}}

\subsection{Definition of $\Gamma$}

\begin{figure}
\centering
\includegraphics[width=\columnwidth]{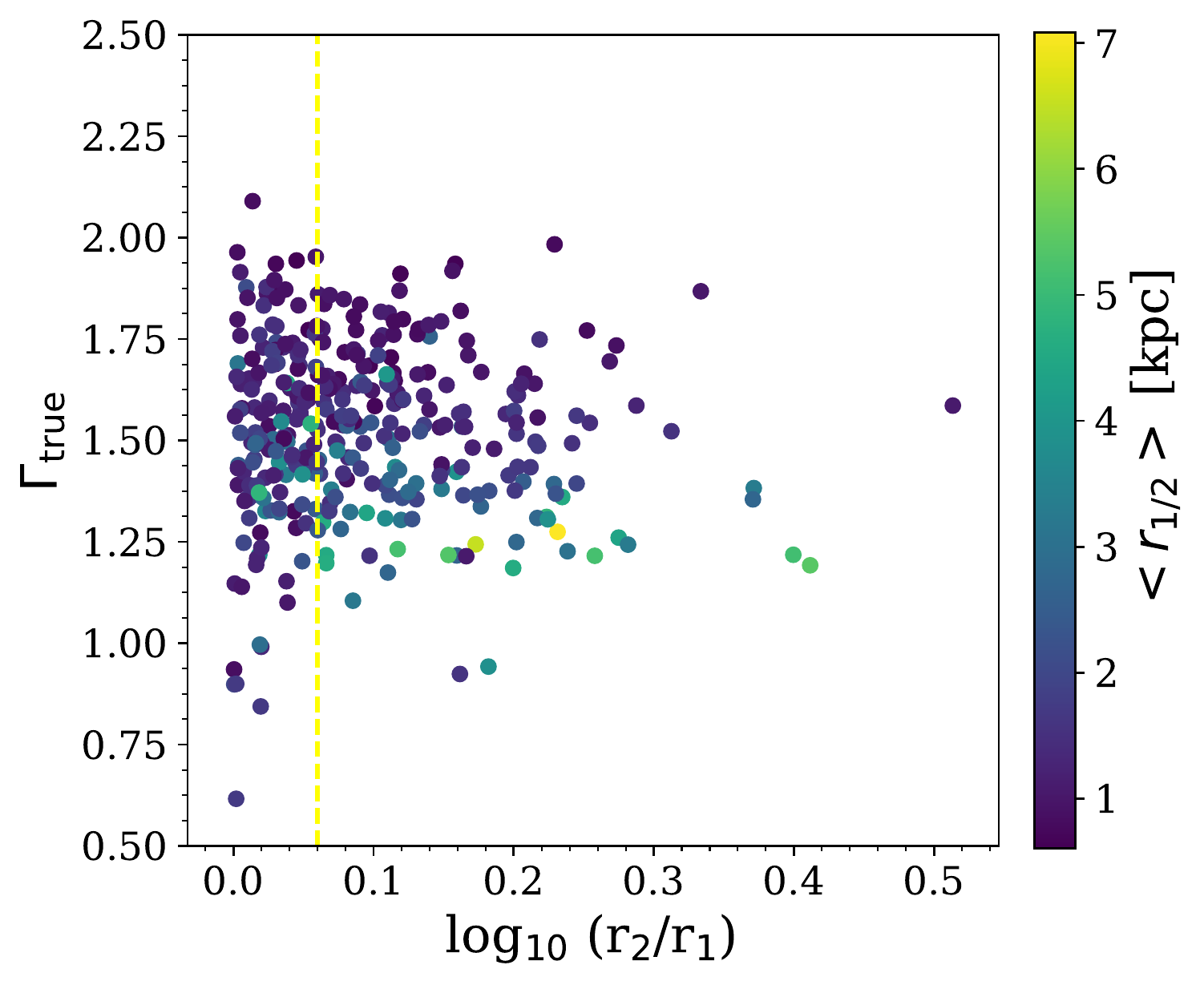}
\caption{True slope of the cumulative mass function as
  a function of the ratio of the half-mass radius of the metal-poor
  and the metal-rich subpopulations. The points are coloured by the
  mean of the metal-poor and the metal-rich half-mass
  radii. $\Gamma_{\mathrm{true}}$ reaches low values below
  $\log_{10}(r_2/r_1)$ $\sim$ 0.06 as a near-zero value of $\Delta
  \log_{10}\rm{M}$ becomes comparable to near zero values of $\Delta \log_{10}
  \rm{r}$. At much larger separations the slope is measured further away from
  the centre and is correspondingly steeper. }
\label{appendix:blowup}
\end{figure}

By definition, the value of $\Gamma$ becomes undefined as the radii of
the two metallicity subpopulations coincide, $\Delta \log_{10} r_2/r_1
\rightarrow 0$. In selecting our sample, it is therefore imporant to
include only galaxies in which the two subpopulations are well
separated. We applied the GMM technique to all galaxies in the five
high-resolution APOSTLE simulations and, for each, we measured the 3D
half-mass radius of the metal-rich and the metal-poor subpopulations
and obtained the slope
$\Gamma_{\mathrm{true}}$. Fig.~\ref{appendix:blowup} shows
$\Gamma_{\mathrm{true}}$ as a function of the ratio of the 3D
half-mass radii of the two metallicity subpopulations.  Below
$\log_{10}(r_2/r_1)$ $\sim$ 0.06, where $r_2$ is the larger radius of
the two, very low values of $\Gamma_{\mathrm{true}}$ are obtained. We require $\log_{10}(r_2/r_1)$~>~0.06 for our sample, as shown by the dashed yellow line in Fig.~\ref{appendix:blowup}.

\subsection{Subpopulation mixing}
\label{mixing}

\begin{figure*}
\centering
\includegraphics[width=2.\columnwidth]{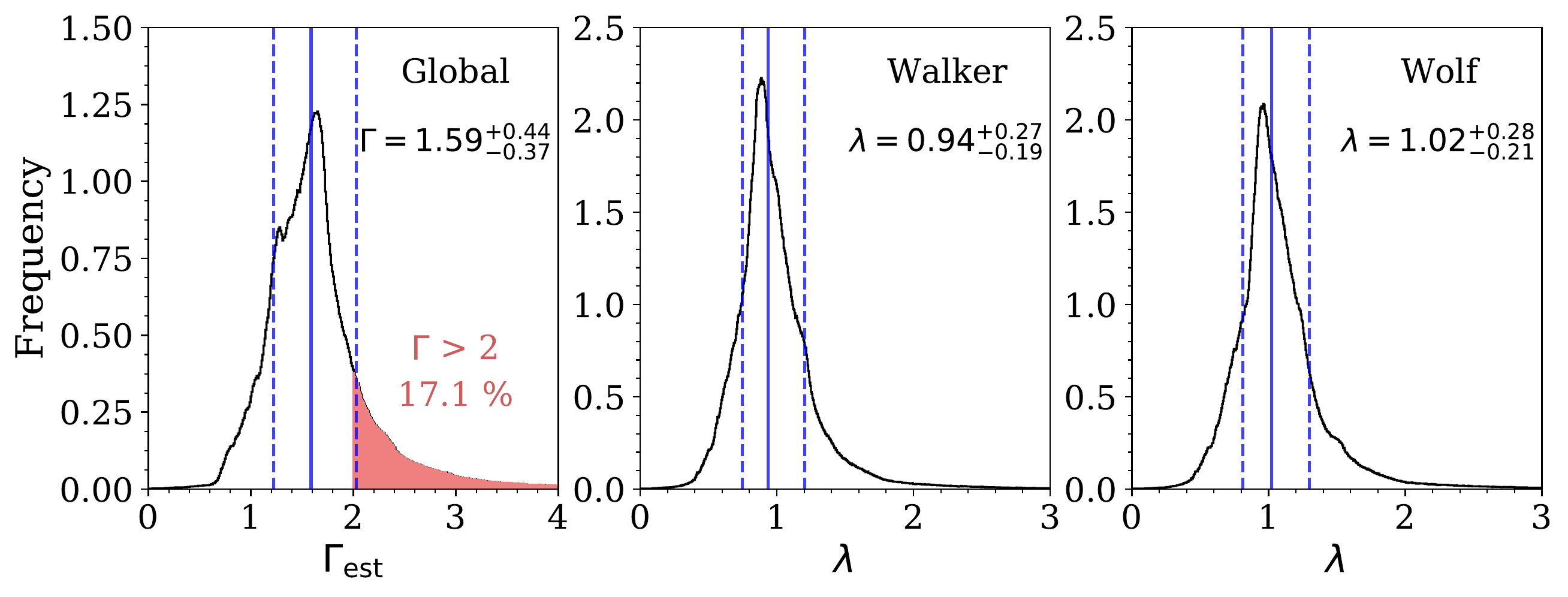}
\caption{Overall distributions of estimated $\Gamma_{\mathrm{est}}$ (left) and $\lambda$ for \citet{walkerest} (middle) and \citet{wolf} (right) estimators for galaxies in our
  sample when a metallicity gap is introduced between the two
  subpopulations, taking out a quarter of each subpopulation on either
  side of the metallicity cut. The distributions are consistent with our previous results
  (top panel of Fig.~\ref{global}), suggesting that kinematic mixing
  between the subpopulations does not play a major role in the
  inferred distribution of slopes $\Gamma$. }
\label{appendix:globalBand}
\end{figure*}

As described in Section~\ref{splitting}, we define each subpopulation
by placing a rigid cut at the intersection of two fitted
Gaussians. This treatment ignores mixing between the two
subpopulations, which can affect the measurement of the velocity
dispersion thus introducing a correlation in the kinematics of the
metal-rich and the metal-poor subpopulations.  \citet{battaglia}, for
instance, allow a metallicity gap of $-1.7 <\left[
  \mathrm{Fe/H}\right]< -1.5$. We therefore repeated our analysis,
this time taking out one quarter of each subpopulation on both sides
of the metallicity cut. The result may be seen in Fig.~\ref{appendix:globalBand}. We obtain $\Gamma$ =
1.59$^{+0.44}_{-0.37}$ and $\lambda$ = 0.94$^{+0.27}_{-0.19}$ for the \citet{walkerest} estimator and $\lambda$ = 1.02$^{+0.28}_{-0.21}$ for the \citet{wolf} estimator when
including the metallicity gap. This result is clearly consistent with
our previous results for the \citet{walkerest} and \citet{wolf} estimators when a metallicity gap was
not included (Section \ref{general}).

\setcounter{figure}{0}
\setcounter{subsection}{0}
\renewcommand*{\thesubsection}{B\arabic{subsection}}
\renewcommand*{\thefigure}{B\arabic{figure}}

\section*{Appendix B}
\label{appendixb}

\subsection{Numerical convergence of galaxy properties}
\label{numerical}
\begin{figure}

\includegraphics[width=\columnwidth]{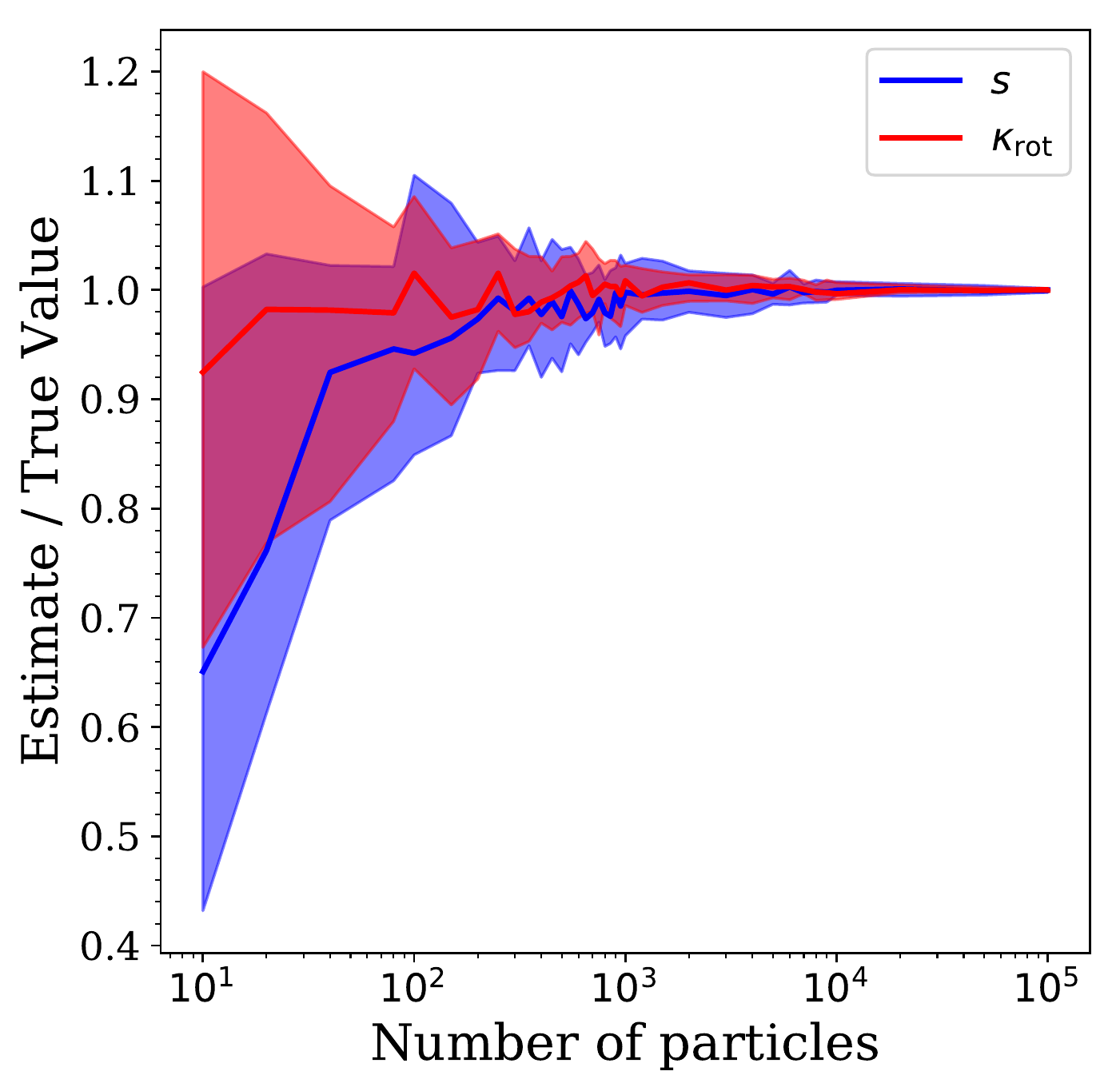}
\caption{The accuracy of the estimates of sphericity $s$ (blue) and $\kappa_{\mathrm{rot}}$ (red) as a function of the number of stellar particles with which they were calculated, shown for a sample of 37 dwarfs containing 10$^5$ to 10$^6$ stellar particles. The `true' values are those calculated using all particles belonging to each galaxy. The blue and red bands represent 1$\sigma$ scatter for $s$ and $\kappa_{\mathrm{rot}}$, respectively. Above a 1000 particles, the values are well converged and at $\sim$400 particles (the minimum number of particles belonging to a subpopulation in our sample of 50 galaxies) the estimates are, on average, accurate and the scatter is less than 10 per cent.}
\label{appendix:convergence}
\end{figure}

As mentioned in Section~\ref{splitting}, we require that the galaxies in our sample include over 1000 stellar particles. Since the stellar particles are then split into two metallicity subpopulations, either or both of these subpopulations may end up with a number of particles too small for the properties such as $\kappa_{\mathrm{rot}}$ and the sphericity~$s$, discussed in Section~\ref{props}, to numerically converge. We therefore carry out the following test for convergence. We take a sample of 37 dwarfs with $10^{5}-10^{6}$ stellar particles from the five high-resolution APOSTLE volumes and we calculate the `true' values of $\kappa_{\mathrm{rot}}$ and $s$, where we include all particles belonging to the galaxy as defined in Section~\ref{splitting}. We then take progressively smaller particle samples and recalculate these properties. The result of this test may be seen in Fig.~\ref{appendix:convergence}, where we show the accuracy of the estimates as a function of the number of particles with which they were calculated. It can be seen that the scatter in the estimate accuracy increases significantly for smaller particle subsamples. The median estimates of $\kappa_\mathrm{rot}$ are generally accurate, even for $<$~100 stellar particles, while the sphericity, $s$, can be substantially underestimated. 

In our sample of 50 dwarfs with two metallicity subpopulations, the minimum number of particles found in an individual subpopulation is just over 400. For this number of particles the estimates are generally accurate, with the 1$\sigma$ scatter of about 5-10 per cent, as can be seen in Fig.~\ref{appendix:convergence}. We thus conclude that the values of $\kappa_{\mathrm{rot}}$ and $s$ presented in this work are not affected by the number of particles assigned to individual metallicity subpopulations to any significant extent.

\label{lastpage}
\end{document}